\documentclass[11pt]{article} 
\usepackage{jheppub}

\usepackage{amsmath,amsfonts,amsbsy,amssymb,array,accents,dsfont}
\usepackage{enumerate}
\usepackage[shortlabels]{enumitem}
\usepackage{graphicx} 
\usepackage{caption} 
\usepackage{subcaption} 
\usepackage{upgreek}
\usepackage{bm}
\usepackage{slashed}

\let\includefigures=\iftrue
\let\useblackboard==\iftrue
\newfam\black

\definecolor{myblue}{RGB}{85,130,255}
\definecolor{myred}{RGB}{200, 45, 40}

\PassOptionsToPackage{ 
     colorlinks=true,
     linkcolor=myBlue,
     urlcolor=blue,
     filecolor=black,
     citecolor=myRed,
}{hyperref}

\usepackage{xparse}
\usepackage{xspace}

\NewDocumentCommand\eqn{om}{%
  \IfNoValueTF{#1}
     {\[ #2 \]}
     {\begin{equation}\label{#1} #2  \end{equation} \expandafter\newcommand\csname #1\endcsname{\eqref{#1}\xspace}\ignorespaces}
}
\NewDocumentCommand\eqna{om}{%
  \IfNoValueTF{#1}
    {\begin{align*} #2 \end{align*}}
    {\begin{equation}\label{#1}\begin{split} #2  \end{split}\end{equation} \expandafter\def\csname #1\endcsname{\eqref{#1}\xspace}\ignorespaces}
}

\newcommand{\rcite}{\cite}

\def\Pt{E}
\def\mhat{{m}}

\def\ptcl{{\rm ptcl}}
\def\str{{\rm str}}
\def\BH{{\rm BH}}

\def\lapse{{ \sfN}}

\def\taub{{\cT}}

\def\sl{\text{sl}}

\def\eps{\epsilon}

\def\ptcl{{\rm ptcl}}
\def\str{{\rm str}}
\def\BH{{\rm BH}}

\def\tight#1{\! #1 \!}  

\def\({\left(}
\def\){\right)}
\def\[{\left[}
\def\]{\right]}

\def\ie{{i.e.}}
\def\eg{{e.g.}}

\def\etc{{etc}}

\def\ext{{\rm ext}}
\def\eff{{\rm eff}}

\def\lstr{\ell_{\textit{s}}}

\def\bh{{\sst\rm BH}}

\def\sfF{{\mathsf F}}

\def\sfH{{\mathsf H}}
\def\sfI{{\mathsf I}}

\def\sfM{{\mathsf M}}
\def\sfN{{\mathsf N}}

\def\sfX{{\mathsf X}}

\def\sfm{{\mathsf m}}
\def\sfn{{\mathsf n}}

\def\sfx{{\mathsf x}}

\DeclareMathSymbol{\medhatsym}{\mathord}{largesymbols}{"62} 

\DeclareMathSymbol{\medtildesym}{\mathord}{largesymbols}{"65}

\makeatletter
\newcommand*\rel@kern[1]{\kern#1\dimexpr\macc@kerna}
\newcommand*\widebar[1]{%
  \begingroup
  \def\mathaccent##1##2{%
    \rel@kern{0.8}%
    \overline{\rel@kern{-0.8}\macc@nucleus\rel@kern{0.2}}%
    \rel@kern{-0.2}%
  }%
  \macc@depth\@ne
  \let\math@bgroup\@empty \let\math@egroup\macc@set@skewchar
  \mathsurround\z@ \frozen@everymath{\mathgroup\macc@group\relax}%
  \macc@set@skewchar\relax
  \let\mathaccentV\macc@nested@a
  \macc@nested@a\relax111{#1}%
  \endgroup
}
\makeatother

\def\half{\frac12}
\def\coeff#1#2{{\textstyle \frac{#1}{#2}}}

\def\Tr{{\rm Tr}}
\def\One{{\hbox{1\kern-1mm l}}}

\def\Im{{\sfI\sfm\,}}

\def\barray{\begin{array}}
\def\earray{\end{array}}
\def\be{\begin{equation}}
\def\ee{\end{equation}}
\def\bea{\begin{eqnarray}}
\def\eea{\end{eqnarray}}
\def\bal{\begin{align}}
\def\eal{\end{align}}
\def\nn{\nonumber}

\newcommand{\bC}{{\mathbb C}}

\newcommand{\bH}{{\mathbb H}}

\newcommand{\bP}{{\mathbb P}}

\newcommand{\bR}{{\mathbb R}}
\newcommand{\bS}{{\mathbb S}}
\newcommand{\bT}{{\mathbb T}}

\definecolor{cardinal}{rgb}{0.6,0,0}
\definecolor{darkgreen}{rgb}{0,0.4,0}
\definecolor{green}{rgb}{0,0.4,0}
\definecolor{golden}{rgb}{0.92, 0.7, 0}
\definecolor{midnight}{rgb}{0, 0, 0.5}
\definecolor{darkblue}{rgb}{0, 0, 0.7}

\numberwithin{equation}{section}

\mathchardef\mhyphen="2D

  \def\cF{\mathcal {F}}
  
  \def\cL{\mathcal {L}}
 \def\cN{\mathcal {N}} \def\cO{\mathcal {O}}
\def\cP{\mathcal {P}}  \def\cR{\mathcal {R}}
\def\cS{\mathcal {S}} \def\cT{\mathcal {T}}

\def\one{{\hbox{\kern+.5mm 1\kern-.8mm l}}}
\def\zero{{\hbox{0\kern-1.5mm 0}}}

\newcommand{\ket}[1]{{\,| {#1} \rangle}}

\def\id{\textrm{id}}

\def\id{{1 \kern-.28em {\rm l}}}

\def\journal#1&#2(#3){\unskip, \sl #1\ \bf #2 \rm(19#3) }
\def\andjournal#1&#2(#3){\sl #1~\bf #2 \rm (19#3) }

\def\ie{{\it i.e.}}
\def\eg{{\it e.g.}}

\def\etc{{\it etc}}

\def\sst{\scriptscriptstyle}

\def\coeff#1#2{{\textstyle{\frac{#1}{ #2}}}}
\def\half{\frac12}

\def\ket#1{|#1\rangle}

\def\One{{1\hskip -3pt {\rm l}}}

\catcode`\@=11
\def\slash#1{\mathord{\mathpalette\c@ncel{#1}}}
\def\eps{\epsilon}

\def\underrel#1\over#2{\mathrel{\mathop{\kern\z@#1}\limits_{#2}}}

\catcode`\@=12

\def\ket#1{\left| #1\right\rangle}

\def\exp{{\rm exp}}

\def\ie{{\it i.e.}}
\def\eg{{\it e.g.}}

\title{
{
\huge The black hole S-matrix in gauge/gravity duality
}}

\author{
Emil J. Martinec
}

\affiliation{
\vskip 0.01cm
Leinweber Institute for Theoretical Physics, Enrico Fermi Institute, and Department of Physics\\ 
University of Chicago,
5640 S. Ellis Ave.,
Chicago IL 60637\\ 
}

\emailAdd{%
e-martinec@uchicago.edu}

\abstract{%
A variety of examples of gauge/gravity duality have a Coulomb branch for the gauge theory dynamics.  We exploit this feature to construct an S-matrix, in which the initial state is a shell of branes converging on the origin of the Coulomb branch.  In the gravitational dual, the shell of branes sources a geometry with a capped throat; the cap descends from the asymptotic region to larger and larger redshift.  A trapped surface forms when the redshift of the cap reaches the point where the excitation of strings stretching between the branes is unsuppressed, and the dual gauge theory deconfines. 
The intermediate state is a long-lived black hole, which then decays via the slow emission of branes back onto the Coulomb branch.  
We compare and contrast the descriptions of this process in the bulk effective field theory and the dual gauge theory; and discuss the consequences of this construction for the black hole information paradox.  

}

\begin{document}
\hypersetup{pageanchor=false}
\begin{titlepage}
\maketitle
\thispagestyle{empty}
\end{titlepage}
\hypersetup{pageanchor=true}
\pagenumbering{arabic}








\section{Introduction and overview}
\label{sec:intro}

There is substantial evidence that in string theory, black holes are simply complicated bound states of the exotic extended objects that are string theory's fundamental constituents.  

A key component of this body of evidence is gauge/gravity duality (for reviews, see~\rcite{Aharony:1999ti,Peet:2000hn}). 
The explicit unitarity of the gauge theory suggests that string theory indeed resolves the various puzzles associated with the quantum theory of black holes, though an understanding of the specific mechanism of this resolution is obstructed by our currently incomplete understanding of the duality map.  
Because the duality relates strong coupling on one side to weak coupling on the other, the brane dynamics is strongly coupled in the regime of interest, and so absent the needed computational tools we are typically left making heuristic arguments and qualitative statements.
This gap has left plenty of room for speculations about candidate features of quantum gravity (islands, wormholes, baby universes, fuzzballs, \etc.) that might supply such a mechanism, but which alternatively might not have any basis in string theoretic reality.

To date, attention has largely focused on situations such as the duality between global $AdS_{d+1}$ spacetime and conformal gauge theory on a sphere $\bS^{d-1}$.  As we will review below, in such a situation, the conformal coupling of the underlying branes to the curvature of the spatial manifold generates a confining potential, and so finite energy processes are localized in the center of $AdS_{d+1}$.

However, to study the black hole information paradox, one is interested in a process whereby an initial black hole is allowed to decay into a collection of Hawking quanta.  A standard approach in the above setup is to extract energy from the system by hand, via smeared local operators in the gauge theory whose action on a thermal state is dual to processes which sample the atmosphere of Hawking quanta surrounding a black hole that is sitting in the middle of the AdS spacetime.  A potential difficulty with this approach is that the analysis of the information paradox then rests on our understanding of the duality map, and precisely what process in the bulk a given (smeared) local perturbation in the gauge theory is representing~-- specifically, where and how it is acting in spacetime.  It has been proposed, for instance~\rcite{Penington:2019npb,Penington:2019kki}, that one can extract Hawking quanta from the bulk and deposit them in some auxiliary system, and then non-locally manipulate the interior state of the black hole by manipulating this auxiliary system without touching the state in the black hole exterior.

If this idea falls into the category of ideas having no basis in string theoretic reality, it might simply be because of our poor understanding of how the holographic map works, and in particular one can worry that the use of local operators in the gauge theory to conduct such manipulations runs the risks of conflating UV and IR effects in the bulk, thereby introducing non-locality where there was none to begin with.

It would be better if one could simply set up a state in the system for which the holographic dictionary is understood, and let it evolve on its own, without the deus ex machina of further manipulation.  But then we need a channel into which any intermediate black hole state can decay.  An obvious candidate is to study situations where the gauge theory has a Coulomb branch in its dynamics.  Instead of a confining box which is the geometric dual to gauge theory on a sphere, we can consider gauge theory on a torus, or any number of other situations where energy can leak out of the system to a far-flung region of the gauge theory configuration space.

One of the early examples of this sort of approach is the BFSS matrix model~\rcite{Banks:1996vh} (for reviews, see~\rcite{Bigatti:1997jy,Banks:1999az,Polchinski:1999br,Taylor:2001vb,Lin:2025iir}).  The decoupling limit of a collection of D0-branes in type IIA string theory is maximally supersymmetric Yang-Mills quantum mechanics; the dual geometry is a region of 11d spacetime (and small excitations within it) described in a highly boosted frame.  In other words, D0-branes carry momentum along the extra circle involved in the lift of IIA string theory to M-theory, and the back-reaction in geometry of a large number $N$ of momentum quanta expands the proper size of that circle to a macroscopic scale in the near-source region where the back-reaction is largest.  The typical bound state of D0-branes having energy of order $1/N$ is a highly boosted 11d Schwarzschild black hole; Hawking radiation is the leakage of sub-clusters of D0-branes onto the Coulomb branch of the gauge theory, along which they escape the vicinity of the remaining black hole~\rcite{Banks:1997tn,Klebanov:1997kv,Horowitz:1997fr,Banks:1997cm,Li:1998ci}.

In the matrix model/M-theory duality map, M-theory compactified on $\bT^p$ is dual to maximally supersymmetric Yang-Mills on the T-dual torus $\widetilde\bT^p$ (again at energies of order $1/N$).  Once again, the decay of thermal states of the gauge theory onto the Coulomb branch is a dual description of the Hawking evaporation of Schwarzschild black holes in M-theory compactified on $\bT^p$.  

Unfortunately, in the strongly coupled regime of the gauge theory where the geometric dual description is the appropriate effective field theory, the D-brane wavefunctions are highly spread out~-- for instance, BFSS estimated the scale of the D0-brane wavefunction to be of order $N^{1/9}$, which is also the size of the region of spacetime in which the curvatures are small and 11d supergravity applies.  It is thus hard to separate the Hawking radiation from the black hole, since their wavefunctions highly overlap.  

However, the spatial extent of the region having a {\it geometrical} description of some sort is far larger~\rcite{Itzhaki:1998dd}.  For instance, for $p=3$ and at large 't Hooft coupling $g_s N$, the geometrical description is valid out to arbitrary radius, asymptoting to the toroidally compactified Poincar\'e section of $AdS_5\times \bS^5$.  

In this work, we consider such situations~-- gauge theories having a Coulomb branch, in which we assemble a black hole via the collapse of a shell of branes, which subsequently decays via radiation of the constituent branes back onto the Coulomb branch.  That such a decay is possible is already established by the matrix theory analysis, and in the situations where the dual geometry is valid out to large radius, the radiated branes become well-separated from  the remaining black hole.  There is thus a black hole S-matrix, transpiring entirely within the context of gauge/gravity duality.  This process is exceedingly slow, as we shall see.  It is however the {\it only} available decay channel in many cases, since the supergravity excitations of the bulk geometry (dual to states created by single-trace operators in the gauge theory) live in bound state wavefunctions confined to the vicinity of the black hole.

Hawking radiation is an essentially thermodynamic process.  As such, the Hawking radiation of a very heavy object such as a brane is highly suppressed, because the first law of black hole mechanics
\be
dS_\BH = \frac{dE - \Phi\,\delta Q}{T_\BH}
\ee
implies that the removal of the significant chunk of free energy required to pull out a rather heavy constituent brane onto the Coulomb branch entails a large reduction in the entropy of the black hole; elementary considerations such as Fermi's Golden Rule, the Eigenstate Thermalization Hypothesis (ETH), detailed balance, \etc, then lead to a very slow decay rate via this channel.  In fact, the decay rate is estimated to be of order~\rcite{Massar:1999wg,Parikh:1999mf,Vanzo:2011wq,Wenren:2017ryk,Martinec:2023plo,Martinec:2023iaf}
\be
\label{radrate1}
\Gamma \propto e^{S_f-S_i} ~,
\ee
as we will review below.

Because every emission is costly, any additional free energy used to give the departing branes a significant radial momentum to take it away from the black hole is also highly suppressed; the Hawking quanta are predominantly produced at or very near threshold.  If the Coulomb branch is a moduli space with a flat potential, then the emitted branes marinate for a long time in the thermal atmosphere of supergravity Hawking quanta, and one might worry that there is plenty of opportunity for there to be some interaction between the black hole and the radiated brane that restores unitarity to the evaporation process.

In order to eliminate this sort of possibility, we can destabilize the Coulomb branch, so that once the brane is radiated it is swept away to the asymptotic region.  One way to destabilize the Coulomb branch is to put the gauge theory on a negatively curved (compact) spatial manifold $\Sigma_p$ rather than a torus.  Now the conformal coupling of the spatial curvature to the gauge theory scalars pushes the branes toward rather than away from the conformal boundary.  This variant of gauge/gravity duality has been considered in~\rcite{Birmingham:1998nr,Emparan:1998he,Emparan:1999gf}, and in particular in the context of black hole formation and evaporation, in~\rcite{Horowitz:2009wm,Barbon:2010us,Wenren:2017ryk}.%
\footnote{A similar instability in a more elaborate setting, namely the Klebanov-Witten solution at finite temperature and chemical potential, has been considered in~\rcite{Henriksson:2019ifu,Henriksson:2021zei}.  Here the throat geometry is asymptotically $AdS_5\times T^{1,1}$, and the instability corresponds once again to nucleating $D3$-branes that discharge the throat.}

In this work, we consider gauge/gravity duality where the conformal boundary of spacetime is either $\bR_t\times \bT^p$ or $\bR_t\times \Sigma_p$ (where $\Sigma_p$ is a compact $p$-dimensional space of constant negative curvature), so that there is a Coulomb branch that is either neutrally stable or unstable.  We investigate the process of formation of an intermediate state black hole via the collapse of a shell of branes from the asymptotic region, self-consistently in an effective bulk description that couples $N$ separate brane effective actions to supergravity.  This effective action is valid when the branes are well-separated, and the stringy excitations that stretch between them are heavy and can be consistently integrated out.  At the point of apparent horizon formation in the gravitational description, these non-abelian excitations become light, and the dual gauge theory deconfines (a dynamical version of the well-known duality between the deconfinement transition in the gauge theory and the Hawking-Page transition in gravity~\rcite{Witten:1998zw}).  

During the collapse phase, the geometry sourced by the shell of branes is a capped throat similar to the bubbling geometries that have been studied in BPS contexts (see~\rcite{Bena:2022rna} for an overview).  In section~\ref{sec:branestars}, we review a recent construction of BPS solutions generated by the sort of framework we have in mind, in which fundamental (F1) strings and NS5-branes are described by worldvolume effective actions coupled to the supergravity effective action.  These solutions include a T-dual description of general {\it superstrata}, which are three-charge (NS5-F1-P) BPS geometries which feature a long $AdS_2$ throat ending in a smooth cap.  The cap arises because the angular $\bS^1\times\bS^3\times \bT^4$ geometry sourced by the branes is supported by the brane fluxes, but when one begins to dive inside the source distribution, Gaussian spheres surround less flux; the cycle threaded by this flux in the angular space shrinks (in this case, the $\bS^3$ threaded by the magnetic 3-form flux $H_3$ sourced by the NS5-branes), eventually smoothly capping off at finite radial depth in the throat.

With this prototype in hand, we move on in section~\ref{sec:D-shells} to the consider toroidally wrapped $Dp$-branes, and specifically D3-branes wrapped on $\bT^3$.  Since much of the D3-brane analysis can be carried out in parallel for D3-branes wrapped on $\bT^3$ and on $\Sigma_3$, we leave the spatial curvature $k$ as a parameter in the more detailed analysis of section~\ref{sec:D3 case}.  We analyze D3-brane motion first in the probe approximation, and then fully back-reacted.  In the latter case, it proves convenient to average the sources over the transverse angular sphere $\bS^5$ and treat the branes as thin domain walls in $AdS_5$.

The picture of the geometry transverse to the D3-branes during the collapse phase is that of a cylindrical throat $\bR\times\bS^5$, capped off smoothly at the radial location of the brane shell.  This shell descends the lengthening throat until the throat is deep enough and the redshift large enough that non-abelian brane excitations are unsuppressed, at which point a trapped surface forms and eventually becomes a black hole.%
\footnote{In the hyperbolic case, since the branes are living in an unstable effective potential, if the shell has insufficient energy, it can bounce off the potential and back out to the asymptotic region without forming an intermediate black hole state.}

We then analyze the rare process of brane Hawking radiation in section~\ref{sec:hawking}, sketching the derivation of the emission rate~\eqref{radrate1}.  We discuss graybody factors, in particular for the hyperbolic case, and show that the typical brane emission takes away a fraction of order $1/N$ of the black hole's total $O(N^2)$ energy, and reduces the entropy by a similar amount.  

We then explore the consequences for the black hole information paradox in section~\ref{sec:unitarity}.


\section{Review of toroidally wrapped branes and black holes}
\label{sec:tori}

Our basic approach follows from recent work using an effective action formalism to describe BPS configurations of toroidally wrapped NS5-branes and fundamental strings that are separated in their transverse space~\rcite{Martinec:2024emf,Martinec:2025npg}.  We begin by reviewing this work, and then describe the analogous approach to the description of D-branes on their Coulomb branch.


\subsection{BPS brane stars and capped throats in the F1-NS5 system}
\label{sec:branestars}

The search for bulk microstructure has led to a rich landscape of BPS solutions of the effective gravity theory such as supertubes (see \eg~\rcite{Skenderis:2008qn} for a review) and superstrata (see \eg~\rcite{Shigemori:2020yuo} for a review), whose $AdS_3/CFT_2$ holographic map is quite precise and well-understood.  

\begin{figure}[ht]
\centering
\includegraphics[width=.7\textwidth]{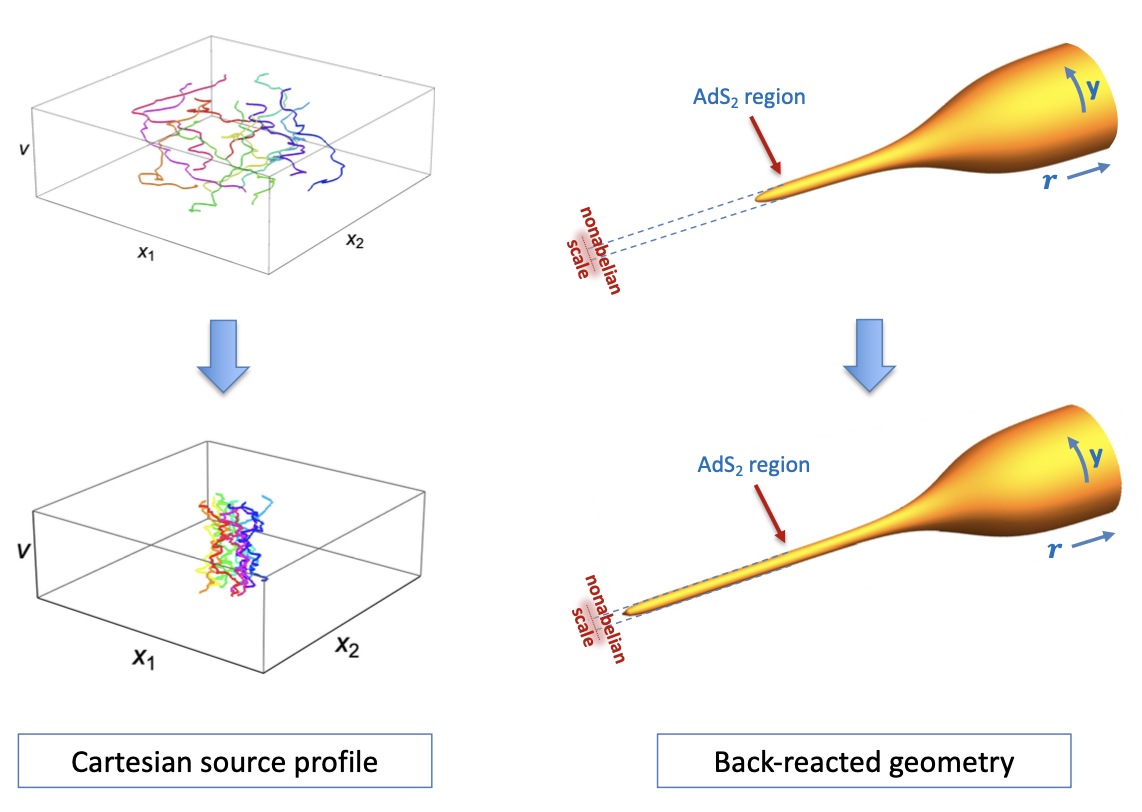}
\caption{\it Slightly separated fivebranes and strings source a capped geometry having an $AdS_2$ throat.  The depth of the throat is related to the spatial separation of the underlying branes.  There is a maximum depth of throat, beyond which perturbative string theory breaks down due to strong coupling in the brane dynamics.}
\label{fig:genST}
\end{figure}

Recently, a large collection of 3-charge BPS bound states of NS5-branes and fundamental (F1) strings carrying arbitrary chiral scalar wave profiles was analyzed using a joint effective action 
\begin{align}
\label{actions}
\cS_{\it sugra} =~& \frac{1}{2\kappa_6^2} \int d^6 x\, e^{-2\Phi}\sqrt{-G} \left( R+ 4 G^{\mu \nu} \partial_{\mu} \Phi \partial_{\nu} \Phi  -\frac{1}{12} H_{\alpha \beta \gamma} H^{\alpha \beta \gamma} \right)~.
\nn\\[.2cm]
\mathcal{S}_{\it branes} = ~& -\sum_{\sfn=1}^{n_1}\bigg[\frac{\tau_{\sst\rm F1}}{2}\int d^2 \sigma \Big(\sqrt{-\gamma} \gamma^{ab} G_{\mu \nu} \partial_a \sfX_\sfn^\mu \partial_{b} \sfX_\sfn^\nu+\epsilon^{ab} B_{\mu \nu} \partial_a \sfX_\sfn^\mu \partial_{b} \sfX_\sfn^\nu  \Big)\bigg]
\\[.2cm]
&-\sum_{\sfm=1}^{n_5}\bigg[\frac{ \tau_{\sst\rm NS5}}{2}\int d^2 \tilde{\sigma} \Big(\sqrt{-\tilde{\gamma}} \tilde{\gamma}^{ab} e^{-2\Phi}G_{\mu \nu} \partial_a \sfF_\sfm^\mu \partial_{b} \sfF_\sfm^\nu+\epsilon^{ab} \widetilde{B}_{\mu \nu} \partial_a \sfF_\sfm^\mu \partial_{b} \sfF_\sfm^\nu  \Big)\bigg]~.
\nn
\end{align}
of bulk supergravity coupled to explicit F1 and NS5-brane sources~\rcite{Martinec:2025npg}.  In this situation, one has precise control over the separation of the branes, and the brane sources can be made self-consistently well-separated so that the solutions to the equations of motion are valid, with non-abelian excitations suppressed.%
\footnote{This approach is well-supported by exact worldsheet descriptions of two-charge (NS5-F1 and NS5-P) backgrounds having exact worldsheet descriptions using gauged Wess-Zumino-Witten models~\rcite{Martinec:2017ztd,Martinec:2019wzw,Martinec:2025npg}, which have a manifestly well-behaved string loop expansion.  While the classical solutions have singularities at the locations of the brane sources, these are artifacts of the supergravity approximation that are completely resolved by $\alpha'$ effects in perturbative string theory.  }
The branes source a capped cylindrical ($AdS_2$) throat, where the depth of the cap correlates to the separation of the brane source distribution in the transverse space, see figure~\ref{fig:genST}. 

A key feature of these generalized superstrata is that the underlying fivebrane sources are slightly separated in their transverse space; it is this separation that is responsible for the $AdS_2$ throat depth being somewhat shallower than that of the quantum black hole carrying the same charges.  A layered, linear structure of the BPS equations of motion leads to a rather direct connection between this slightly ``Coulomb branch'' source configuration and the fully back-reacted geometry, as depicted in figure~\ref{fig:genST}.

Scaling the separation of the sources scales the radial position of the source distribution.  It was seen that moving along the BPS configuration space by scaling the transverse extent of the source distribution, the scale at which non-abelian degrees of freedom begin to be excited matches the scale where quantum fluctuations of the near-horizon region become large as one lowers the temperature of the corresponding three-charge black hole to extremality~\rcite{Preskill:1991tb,Lin:2022zxd}.  But until that regime is reached, the non-abelian excitations are heavy and can be self-consistently integrated out, leaving a low-energy effective field theory~\eqref{actions} of separated branes coupled to supergravity, whose solutions describe perturbative string backgrounds.

%
%
%
%

If one compresses the source configuration by some factor $\lambda$, the depth of the throat increases correspondingly.  When the depth of the throat reaches the ``non-abelian scale'' where the NS5-branes are close enough that non-abelian excitations are liberated, a horizon forms.  In examples that have been worked out~\rcite{Bena:2025uyg,Martinec:2025npg}, the depth of the $AdS_2$ throat at the non-abelian scale is approximately $R_{\sst AdS2}\log[S_\ext]$ where $S_\ext = 2\pi\sqrt{n_5n_1n_p^{\vphantom{a}}}$ is the extremal entropy.  Not coincidentally, this is also the depth in the throat where quantum fluctuations of the horizon of non-extremal $AdS_2$ black holes become large~\rcite{Preskill:1991tb,Lin:2022zxd,Kolanowski:2024zrq,Martinec:2025npg}.

In the near-BPS context, one can consider dynamical processes where the brane star adiabatically collapses to make a near-extremal black hole.  The brane matter forms the cap of the geometry, and as this shell of matter compresses, the $AdS_2$ throat deepens more and more until the non-abelian fivebrane dynamics deconfines, and a near-extremal BTZ black hole forms.  The non-abelian excitations of NS5-branes are {\it little strings}~\rcite{Dijkgraaf:1997ku,Seiberg:1997zk,Maldacena:1996ya}~-- when $n_5$ fivebranes coincide, they can trap fundamental strings by fractionating them into $n_5$ constituent little strings.  Black fivebrane thermodynamics is the Hagedorn entropy of little strings~\rcite{Maldacena:1996ya}; restricting that Hagedorn entropy to the superselection sector of large little string winding and taking the $AdS_3$ decoupling limit, the Hagedorn entropy reduces to the BTZ/Cardy entropy~\rcite{Martinec:2019wzw}:
\begin{align}
S_{ \it little} &= 2\pi\Big(\sqrt{N_L} + \sqrt{N_R}\,\Big)
~\longrightarrow~ S_{\rm BTZ} = 2\pi\Big( \sqrt{n_5n_1(\varepsilon + n_p)/2} + \sqrt{n_5n_1(\varepsilon - n_p)/2}\, \Big)
\end{align}
But little string deconfinement does not happen until the fivebranes are sufficiently close to one another, and this doesn't generically happen until the throat is sufficiently deep.

Thus, one can form a black hole by collapsing a collection of fivebranes and strings.  The separation of the sources controls the depth of a capped throat, and when that throat becomes sufficiently deep, an apparent horizon forms.  The moniker ``apparent'' is appropriate, because it is believed that there are no actual horizons in string theory in the sense of a causal barrier surrounding a black hole.

In the $AdS_3$ decoupling limit, this black hole cannot typically evaporate via the emission of ordinary supergraviton Hawking quanta, which lie in bound states in a thermal atmosphere that is confined to the vicinity of the black hole by the ambient $AdS_3$ geometry.  The fundamental strings that make up the black hole {\it can} however reach the $AdS_3$ boundary at finite cost in energy~\rcite{Seiberg:1999xz,Maldacena:2000hw}, and have a finite probability to be Hawking radiated~\rcite{Martinec:2023plo,Martinec:2023iaf}.  The emission probability is (taking into account the chemical potential for charge emission, and measuring energy relative to the appropriate production threshold)%
\footnote{A similar analysis of the emission of charged particles from a charged black hole was undertaken in~\rcite{Brown:2024ajk}.}
\be
\label{Gambh}
\Gamma \sim e^{\delta S_{\BH}} = e^{S_f-S_i} ~.
\ee
which for fundamental string emission is
\be
\delta S \sim \frac{\delta E - \Omega\, \delta P}T - \pi n_5 r_+ ~,
\ee
where $\Omega=r_-/r_+$, and $r_\pm$ are the inner and outer horizon radii.  This result is further corrected by graybody factors, which we will estimate for a D3-brane example below.  In general, one expects such factors to make an order one change to the coefficient of the exponent in the emission probability, but not to change the parametric dependence on the charges, \etc.

Thus, even though one is working in $AdS_3$ spacetime, black holes have an S-matrix, that proceeds by collapsing the brane constituents, and then very slowly re-radiating them back onto their Coulomb branch.    

In this work, we will analyze analogous phenomena in the context of $AdS_5$ black holes, and discuss their consequences for the black hole information paradox and the various ideas that have been proposed to address it.  We begin with a general discussion of gauge/gravity duality for toroidally wrapped $Dp$-branes, which all have a Coulomb branch moduli space of configurations.


\subsection{Toroidally wrapped Dp-branes}
\label{sec:D-shells}

Another near-BPS context where gauge/gravity duality leads to a black hole S-matrix is the BFSS matrix model~\rcite{Banks:1996vh}.  There, a black hole in M-theory compactified on $\bT^p$ ($p=0,1,2,3,4$) is described as a bound state of $N$ toroidally compactified {\it Dp}-branes, at energy scales of order $1/N$.

At somewhat larger scales, the geometry sourced by such toroidally wrapped black {\it Dp}-branes in the gauge/gravity duality decoupling limit is~\rcite{Itzhaki:1998dd,Martinec:1998ja}:
\begin{align}
\label{IMSY}
ds^2 &= H^{-1/2}\Big(-f(r) dt^2 + ds^2_{\bT^p}\Big) + H^{1/2} \Big(\frac{dr^2}{f(r)} + r^2 d\Omega_{8-p}^2 \Big)
\nn\\[.3cm]
f(r) &= 1-\Big(\frac{r_0}{r}\Big)^{7-p}
~~,~~~~
H(r) = \Big(\frac{r_p}{r}\Big)^{7-p}
~~,~~~~
r_p^{7-p} = d_p (\alpha')^{\frac{7-p}2}g_s N
\\[.3cm]
e^{\Phi} &= g_s H^{\frac{3-p}{4}}
~~,~~~~
C_{01...p} =  g_s^{-1}\half\big( H^{-1} -1 \big)  ~,
\nn
\end{align}
where $d_p=(4\pi)^{\frac{5-p}{2}}\Gamma\big(\coeff{7-p}2\big)$.

Again, it will be important for us that the brane dynamics has a Coulomb branch.  BPS Coulomb branch configurations set the non-extremality $r_0$ to zero, and replace the harmonic function $H$ by its multicenter counterpart
\be
\label{HCoul}
H(\sfx) = \frac1N\sum_{\sfn=1}^N \Big(\frac{r_p}{|\sfx-\sfx_\sfn|} \Big)^{7-p}  ~.
\ee
In the BPS context, the $\sfx_\sfn$ have a flat potential, and so can be freely chosen.%
\footnote{Technically, the branes are wrapped on a compact space and so the scalars don't have vev's, rather there is some zero mode wavefunction.  So one should make localized wavepackets and dispense with this red herring.}
The gravitational attraction of the branes exactly balances electrostatic repulsion of the Ramond-Ramond gauge field $C_{p+1}$.  When the system is excited above the BPS bound, the two are no longer exactly in balance, but the effective potential is asymptotically flat at large brane separation.

The geometry of a shell of branes is again a capped throat, for the same reason~-- Gaussian spheres at radii $r=|\sfx|\gg|\sfx_\sfn|$ that enclose all $\sfn$ branes, enclose the flux that supports the $\bS^{8-p}$ at a radius approximately $r({r_p}/{r})^{\frac{7-p}{4}}$.  The scale $r_p$ depends on $N$, and so the sphere shrinks away smoothly in the core of the shell as it surrounds fewer and fewer branes. 
One can then collapse the shell by giving it some radial kinetic energy, leading to the black brane geometry~\eqref{IMSY}.

Note that asymptotic boundary in the bulk dynamics is always the vacuum $Dp$-brane geometry, with the conformal boundary being $\bR_t\times \bT^p$.%
\footnote{More precisely, for $p<3$ the spacetime curvature grows toward the conformal boundary, and the local string coupling weakens; beyond some large radius, the geometrical description becomes strongly coupled, and the gauge theory description is weakly coupled. }
The UV field theory modes are always in their ground state, correspondingly there are also no supergravity excitations in the bulk geometry at asymptotically large radius.  There are always modes on scales arbitrarily large relative to the Coulomb vevs at any given time, but such modes are not excited.  On scales below that of the vev, there are only the finite number of ground states of the individual branes, and their collective center of mass motions, which constitute the microstructure of the cap in the geometry; otherwise there is nothing describing a region of spacetime that is ``more infrared'' than the scale of the vevs.  The absence of excitations in the gauge theory below some energy scale (other than the collective motions of the branes) is dual to the existence of a cap in the geometry.   But the radial position of the cap can scale up or down easily by dialing the scale of the vevs.

The idea of BFSS is that as one lowers the temperature so that the horizon radius $r_0$ decreases, the proper size of the torus wrapped by the branes is shrinking in the near-horizon region, and eventually becomes smaller than the string scale.  At this point, the appropriate duality frame for the description of the near-horizon physics is the T-dual one; the $Dp$-branes T-dualize to D0-branes, and one is describing D0-branes moving around in the T-dual torus as well as the rest of the T-dual geometry.  As one continues to lower the temperature, the dilaton at the horizon scale of this T-dual geometry grows, and eventually the near-horizon region lifts to M-theory.  Geometrically, D0-branes are simply momentum sources in the extra circle of M-theory; their back reaction expands that extra circle near the brane source, creating a little bubble of 11d spacetime in which highly boosted objects are interacting~\rcite{Horowitz:1997fr,Martinec:1998ja,Martinec:1999sa}.

In the dual gauge theory, as the temperature is correspondingly lowered, eventually the available free energy is unable to excite all but the zero modes of the gauge theory, and one arrives at an effective quantum mechanics of the zero-mode sector of the gauge theory, which is identical to the dynamics of the T-dual D0-branes~\rcite{Banks:1996vh,Taylor:1996ik}.  

One can see {\it qualitative} features of 11d black holes arising from the assembly of separated D0-branes as one brings them together.  For instance, a combination of the virial theorem and the uncertainty principle correctly reproduces the Schwarzschild radius as the approximate size of the bound state~\rcite{Horowitz:1997fr,Li:1998ci}, and Hawking radiation is described in this highly boosted frame as the process of radiating D0-branes (or D0-brane threshold bound states) back onto the Coulomb branch of the gauge theory~\rcite{Banks:1997tn,Horowitz:1997fr,Banks:1997hz,Li:1997zf}.

The difficulty of making progress in this setup is that all the important features of the bulk microstructure are clouded by the fact that zero-mode wavefunctions of the branes in the strongly-coupled brane dynamics are spread out over scales much larger than those of the objects under study.  For instance, the extent of the D0-brane wavefunction in the original BFSS construction is estimated to be of order $N^{1/9}$, parametrically much larger than the scale $r_0$ of the M-theory black hole that is being described as a metastable thermal bound state of the D0-branes with energy of order $1/N$.  One must somehow tease out of this diffuse cloud of branes the bulk microstructure.
However, we don't need to focus solely on the ultra-low temperature regime of the gauge theory dynamics considered by BFSS; the geometry~\eqref{IMSY} is valid out to much larger radius~\rcite{Itzhaki:1998dd,Martinec:1998ja,Martinec:1999sa}.

Generalizing the BFSS analysis, we can consider the formation of toroidal {\it Dp}-brane black holes by assembling them from initial states of branes well-separated on their Coulomb branch, followed by their slow evaporation back onto the Coulomb branch~\rcite{Martinec:2023iaf}. 
The emission of individual Dp-branes onto the Coulomb branch of the gauge theory proceeds by a tunneling calculation (using techniques pioneered in~\rcite{Kraus:1994by,Kraus:1994fj,Parikh:1999mf}; see section~\ref{sec:hawking} below), leading to the result~\eqref{Gambh}~\rcite{Wenren:2017ryk,Martinec:2023plo,Martinec:2023iaf} that the emission probability is exponentially suppressed by the change in entropy.  

Such brane emissions, governed by the change in entropy according to~\eqref{radrate1}, are quite rare.
For D3-branes, the black hole entropy is of order $N^2$, so with the emission of a single D3-brane the entropy changes by an amount of order $N$, and emission is heavily suppressed.%
\footnote{The equation of state in the gravity regime for general $p$ is
\be
\big(\lstr E\big)^{9-p} = g_s^{3-p} \Big(\frac{R}{\lstr}\Big)^{p(p-5)} \Big(\frac{S^2}{N}\Big)^{7-p} 
\ee
where $R$ is the cycle size of the torus.  From this one deduces
\be
\frac{\partial S}{\partial N} \sim g_s^{\frac{-(9-p)(3-p)}{5-p}}N^{\frac{2}{5-p}}\Big(\frac{R}{\lstr}\Big)^p \Big(\lstr T\Big)^{\frac{9-p}{5-p}} ~.
\ee
Thus the emission is always suppressed by a power of $N$.
}
Once emitted, a brane can run back out on the Coulomb branch and escape the system.
But as mentioned in the Introduction above, an issue with this scenario is that the typical emitted brane is produced very near threshold, as any additional energy used to supply the brane with radial momentum to take it away from the black hole costs extra free energy, and therefore entropy, and is therefore thermodynamically suppressed.  One would like to arrange a clean separation between the radiated quanta and the black hole left behind, so that there can be no claim that information is somehow being transferred to the radiated brane during the long period of time it spends in the thermal atmosphere of the black hole before finally leaving its vicinity.

One can engineer such behavior by destabilizing the Coulomb branch.  There are a variety of options for doing so; a simple one is to replace the toroidal space wrapped by the branes for a compact manifold of constant negative curvature.  We turn now to a discussion of the dynamics of branes wrapped on such hyperbolic spaces, following~\rcite{Horowitz:2009wm,Barbon:2010us,Wenren:2017ryk}.  Since the curvature $k$ of the spatial geometry enters quite trivially as a parameter in the metric coefficients, we can and will treat the cases of toroidal and hyperbolic branes in parallel in what follows, focussing on the conformal example with $p=3$.


\section{D3-branes on compact manifolds}
\label{sec:D3 case}

Consider the black D3-brane geometry in the decoupling limit; this will closely approximate the intermediate state after the initial collapse process
\begin{align}
\label{D3geom}
ds^2 &= \ell^2\Big[-f(r) dt^2 + r^2 d\sfM_3^2\Big]_\parallel + \ell^2 \Big[\frac{dr^2}{f(r)} + d\Omega_5^2 \Big]_\perp
\nn\\[.1cm]
f(r) &= r^2 + k - \frac{\mu}{r^2}  
~~,~~~~
\ell^4 = \frac{g_s N \ell_s^4}{4\pi^3}
\\[.3cm]
C_4 &= \ell^4 r^4 ~;
\nn
\end{align}
here $\ell$ is the $AdS$ scale ($\ell=r_3$ in~\eqref{IMSY}), 
and the subscripts in the metric factors denote directions parallel and transverse to the branes.  The coordinates here are dimensionless.

We allow the spatial manifold $\sfM_3$ wrapped by the D3-branes to be any compact manifold of constant curvature $k$.   
There are three options to choose from~\rcite{Birmingham:1998nr}:
\begin{enumerate}
\item
Put the gauge theory on a spatial sphere $\sfM_3=\bS^3$, so that $k=+1$.  The conformal coupling of scalars in the gauge theory provides a confining potential which lifts the Coulomb branch~-- the geometry is asymptotically that of global $AdS_5$.  
Left to its own devices, a large black hole will not decay.  As discussed above, imprecision in understanding the duality map has led to confusion about what happens when energy is extracted from the system via perturbations at/near the conformal boundary.  
We will thus not consider this case further.
\item
Alternatively, one can put the gauge theory on a torus $\sfM_3=\bT^3$ as discussed above, in which case $k=0$.  Supergravity perturbations are still confined, but the gauge theory has a Coulomb branch moduli space, and a thermal state of the gauge theory can decay spontaneously via the emission of branes, as we review below.  This radiation of the underlying branes is exponentially (thermally) suppressed at high temperature~\rcite{Martinec:2023iaf} because the emission of even a single brane results in a large reduction in the entropy of the black hole left behind, and so is strongly disfavored thermodynamically.
\item
Finally, one can consider gauge theory on a compact hyperbolic space $\sfM_3=\bH_3/\Gamma\equiv \Sigma_3$ (with $\bH_3$ the Euclidean hyperbolic 3-ball and $\Gamma\subset SL(2,\bC)$ a suitable Kleinian group) also has a Coulomb branch, but now the conformal coupling of the gauge theory scalars to the spatial curvature generates an instability whereby branes want to run off to the conformal boundary~\rcite{Horowitz:2009wm}.  Once again, radiation of the underlying branes from such a hyperbolic black hole is exponentially (thermally) suppressed at high temperature.  However, in this case, once produced, the brane is rapidly swept away to the asymptotic region by the unstable potential, and so once it gets past the graybody potential of the near-horizon region, the brane quickly leaves the thermal atmosphere of the black hole and decouples.
\end{enumerate} 
The lapse function $f(r)$ in~\eqref{D3geom} vanishes at the horizon radii
\be
\label{horrad}
r^2_\pm = \half\big( -k \pm \sqrt{k^2+4\mu}\,\big) ~;
\ee
$r_-$ is the location of an inner horizon when it is real.
For the toroidal case, $k=0$, black holes have positive mass $\mu>0$ due to the BPS bound; there is no inner horizon, and the singularity at $r=0$ is always spacelike.  In the hyperbolic case, $k=-1$, the black hole solution exists for $\mu>-\frac14$; in the range $-\frac14<\mu<0$, there are both inner and outer horizons, and a timelike singularity at $r=0$, while for $\mu>0$ there is again no inner horizon ($r_-$ is purely imaginary), and the singularity is again spacelike.

The BPS bound of the toroidal case precludes the consideration of negative mass solutions altogether.  On the other hand, there is no reason to exclude $\mu<-\frac14$ from the spectrum for $k=-1$; there is no supersymmetry for the gauge theory on a hyperbolic spatial geometry, and we will see that backgrounds with large negative mass have physical meaning.  The energy landscape is unbounded below, and the system has no vacuum state.  Nevertheless, like the D0-brane quantum mechanics that describes non-perturbative 1+1d non-critical string theory (see~\rcite{Klebanov:1991qa,Ginsparg:1993is,Martinec:2004td} for reviews), we expect that there is a consistent quantum theory of 3+1d $\cN=4$ super Yang-Mills on compact hyperbolic three-manifolds $\Sigma_3$, whose strong-coupling limit is described by~\eqref{D3geom}.


\subsection{D3-brane dynamics -- Probe approximation}
\label{sec:D3}

To begin, we consider the motion of a probe D3-brane in the black brane background~\eqref{D3geom}.  We then approximate the spacetime as being sourced by a shell of branes, whose trajectory is self-consistently determined by the probe analysis.  A proper treatment, including back-reaction of the brane shell, is undertaken in the next subsection.

In analyzing {\it Dp}-brane dynamics, it is convenient to introduce an auxiliary metric $\gamma_{ab}$ on the brane worldvolume via~\rcite{AbouZeid:1997mt}
\be
\label{linact}
\cS_{Dp} = -\half \tau'_p\int\! d^{p\tight+1}\!\xi\, e^{-\Phi^*}\sqrt{-\gamma}\,\Big[ \gamma^{ab}G^*_{ab} - (p-1)\Lambda\Big] - \tau_p\int\!d^{p\tight+1}\!\xi \,C^*_{p+1}  ~,
\ee
where 
\be
\label{taup}
\tau_p' = \Lambda^{\frac{p-1}{2}} \tau_p ~,
\ee
and star denotes pullback to the brane worldvolume.
For our present purposes, we can ignore the complications in the brane effective action arising from the worldvolume gauge field and the pullback of the NS $B$-field; they will not be important participants in the abelianized dynamics we are considering.%
\footnote{The gauge field is of course a crucial participant in the deconfined phase that describes black D3-branes; however, we will take the initial state of a collapsing shell of branes to have the abelian gauge fields on the separated branes unexcited.}
The metric $\gamma_{ab}$ simplifies the analysis of worldvolume coordinate invariance.  Eliminating it via its algebraic equation of motion 
\be
\label{gameom}
G_{ab}^* = \gamma_{ab}\,\Lambda
\ee
takes us back to the DBI action~\eqref{linact}.  The scale $\Lambda$ is completely arbitrary, and can be set to unity by rescaling all the worldvolume coordinates via $\xi\to \Lambda^{-1/2} \xi$.
We will use the spatial components of the $\gamma_{ij}$ equation of motion to determine the spatial volume element, which for $p=3$ in the coordinates~\eqref{D3geom} is
\be
\label{sqrtgam}
\sqrt{\gamma^{~}_{\it spatial}} = \frac{R^3}{\Lambda^{3/2}} 
\ee
(more generally, for $Dp$-branes one has $[\Lambda^2 H(R)]^{-\frac p4}$).
Also, the equations of motion for the spatial coordinates $X^i(\xi)$ of the brane wrapping $\sfM_p$ are solved by $X^i =\xi^a \delta_a^i$. 

On the other hand, as one typically does for $p=1$, one can fix the worldvolume coordinate invariance via the gauge choice
\be
\label{tempgauge}
\gamma_{0a} = -\delta_{0a}
\ee
while imposing the worldvolume Hamiltonian and momentum constraints.  The Hamiltonian constraint sets
\begin{align}
0 &= \partial_0 X^\mu \partial_0 X^\nu G_{\mu\nu}(X) 
+ \Lambda 
\nn\\
&= \ell^2\Big(-\dot T^2 f(R) + \frac{\dot R^2}{f(R)}\Big) + \Lambda
~,
\label{hamcon}
\end{align}
where we have used the spatial components of~\eqref{gameom}.
Here and below, dots refer to derivatives with respect to worldvolume time, which in this gauge (upon rescaling $\xi_0$ to set $\Lambda=1$) is proper time.

In addition, time translation symmetry of the background geometry~\eqref{D3geom} yields the conserved Killing momentum
\be
\label{Pt}
\Pt = -\frac{\delta \cS_{Dp}}{\delta \dot T} = \tau'_p e^{-\Phi} \sqrt{-\gamma}\, G^*_{00}\,\dot T + \tau_p C_{0i_1...i_p}\,\partial_{a_1}\!X^{i_1}...\partial_{a_p}\!X^{i_p} ~.
\ee
 
One can then solve for $\dot T$;
substituting into the Hamiltonian constraint yields (setting $p=3$, $\Phi=0$, and rescaling the coordinates to set $\Lambda=1$) 
\begin{align}
\label{vels}
\begin{split}
\dot T &=  \frac{\Pt+m R^4}{m  R^3 f(R)}
\\[.2cm]
\dot R^2 &= 
\frac{(\Pt+
m R^4)^2}{m^2 R^{6}} -   f(R) ~, 
\end{split}
\end{align}
where
\be
\label{mdef}
m = \tau_3\ell^4 V_{\sst\sfM} = \frac{ N V_{\sst\sfM} }{2\pi^2} 
\ee
is an effective particle mass obtained from dimensional reduction along the brane worldvolume.

The leading term at large $R$ in the radial velocity cancels between the electrostatic repulsion arising from the four-form potential in the Chern-Simons term and the gravitational attraction arising from the DBI term.  The large $R$ behavior is controlled by the subleading term  
\begin{align}
\label{velsolDBI}
\dot R^2 ~&\sim~ -k +\frac{2E}{m R^2} +\dots 
\nn\\[.3cm]
\Big(\frac{dR}{dT}\Big)^2 &\sim -k R^2 +\frac{m(\mu-2k^2)+2E}{m} +\dots
\end{align}
For toroidally wrapped branes, $k=0$, 
the velocity is approximately constant at large $R$; the motion has no turning points, and describes either a probe brane that is falling in toward the black hole, or a radiated Hawking quantum that is headed out toward the asymptotic region. 

For the hyperbolic case $k=-1$, the velocity is asymptotically proportional to the radius, characteristic of motion in an inverted quadratic potential.
A probe brane with sufficiently negative energy has a turning point at a radius $R_*$, where
\be
\label{turnaround}
{R_*^2} \sim -\big(\mu-2\big) - \frac{2E}{m} 
\ee
(note that $\Pt<0$ in order for the probe brane to reflect off the potential barrier).

\begin{figure}[ht]
\centering
\includegraphics[width=.25\textwidth]{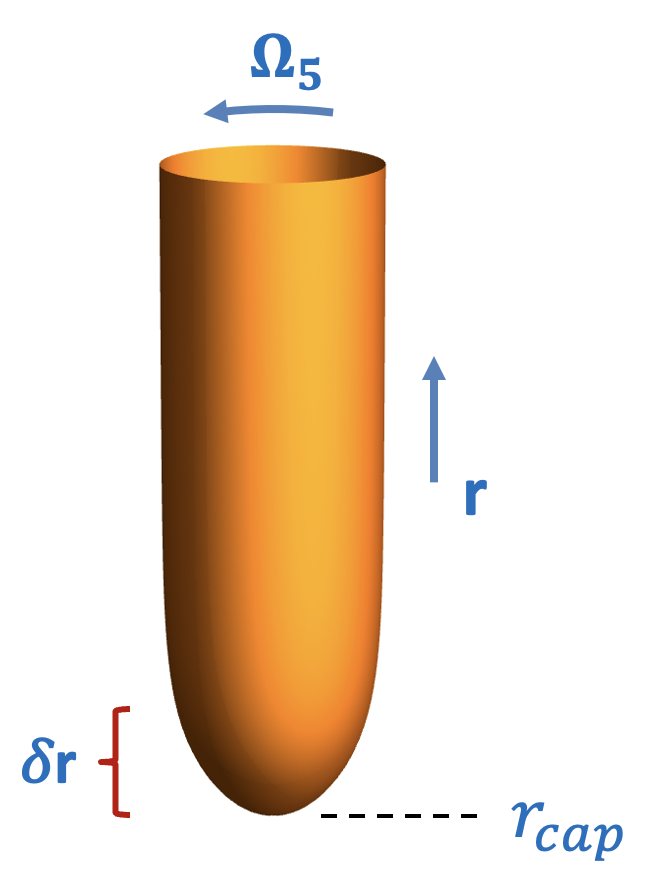}
\caption{\it Snapshot at fixed time of the capped throat transverse geometry sourced by a shell of branes.}
\label{fig:CappedThroat}
\end{figure}

Rather than considering a single brane in the probe approximation moving in a black hole geometry that has already formed from a prior collapse, we would like to consider the collapse process itself.  A self-consistent approximation scheme treats the motion of each brane in the background sourced by the other branes in the shell, for which the probe analysis above is a reliable guide.
After a shell of branes has passed inside some radius in the throat, the geometry outside the shell is the vacuum geometry~\eqref{D3geom} to a good approximation.  A Gaussian sphere at large radius measures all the magnetic five-form flux sourced by the D3's, and so the radius of $\bS^5$ is given by
\be
\ell^4 = \frac{Ng_s\lstr^4}{4\pi^3} ~.
\ee
As one brings the Gaussian sphere into the shell of branes, it encloses less and less flux, and so the geometry sourced by the branes inside the shell is smaller.  Moving steadily into the core of the shell, at its inner edge the size of the $\bS^5$ smoothly shrinks to zero, and the geometry caps off.  The back-reacted geometry sourced by the brane shell will be a capped throat, see figure~\ref{fig:CappedThroat}, much like one finds in the AdS bubbling geometries studied in the literature, and in the present context mirrors the $AdS_2/AdS_3$ examples depicted in figure~\ref{fig:genST}.  We now turn to an analysis of the back-reaction of these brane shells.

%
%
%
%

\subsection{D3-brane shells -- Coupling to gravity}
\label{sec:gravity}

Following the approach of~\rcite{Martinec:2024emf,Martinec:2025npg}, we will treat the dynamics of a brane shell as a sum over $N$ isolated D3-branes on their Coulomb branch, where the effective action of each individual D3-brane is the DBI action~\eqref{linact}, coupled to supergravity
\begin{align}
\label{jointeffact}
\cS_{\it sugra} &= \frac{1}{2\kappa^2} \int d^6 x\, \sqrt{-G}\bigg[ e^{-2\Phi} \Big( R+ 4 G^{\mu \nu} \partial_{\mu} \Phi \partial_{\nu} \Phi  -\frac{1}{2} \big|H_3\big|^2 \Big) 
\nn\\[.2cm]
&\hskip 3.5cm
-\half \Big( \big|F_1\big|^2 + \big|F_3\big|^2 +\half \big|F_5\big|^2\Big)\bigg] - \frac{1}{4\kappa^2} \int C_4\wedge H_3\wedge F_3  
\nn\\[.3cm]
\cS_{Dp} &= \sum_{\sfn=1}^N \bigg[-\half \tau'_p\int\! d^{p\tight+1}\!\xi\, e^{-\Phi}\sqrt{-\gamma}\,\Big( \gamma^{ab}G_{\mu\nu}(\sfX_\sfn)\partial_a\sfX_\sfn^\mu\partial_b\sfX_\sfn^\nu - (p-1)\Lambda\Big) 
\\[.2cm]
&\hskip 1.6cm
- \tau_p\int\!d^{p\tight+1}\!\xi \,C_{\mu_0\cdots\mu_p}(\sfX_\sfn)\,\partial\sfX_\sfn^{\mu_0}\wedge\cdots\wedge\partial\sfX_\sfn^{\mu_p} \bigg]  ~,
\nn
\end{align}
where $\tau'_p$ was defined in~\eqref{taup}, and implicitly $p=3$.
This effective action treats the collection of D3-branes as $N$ isolated sources, suppressing non-abelian dynamics in the effective action.  This ansatz is self-consistent if the branes are far enough out on their Coulomb branch that the non-abelian degrees of freedom~-- the strings stretching between different D3's~-- are sufficiently heavy that they can be integrated out. The string action is (ignoring the coupling to the R-R background, which plays no central role here)
\be
\label{Sstr}
\cS_{\rm\sst F1} = \frac{1}{4\pi \alpha'}\int \! d^2\sigma  \sqrt{\gamma} \gamma^{ab} \partial_a X^\mu \partial_b X^\nu G_{\mu\nu}(X)
\ee
In the background~\eqref{D3geom}, a string stretched between branes separated by an angle $\delta\Omega_5$ on $\bS^5$ has an energy 
\be
\label{Estr}
E_\str \approx \frac{\ell}{\alpha'} \sqrt{f(r)}\,\delta\Omega_5
\ee
where the factor of the redshift factor $f(r)\sim r^2$ appears via the solution of the reparametrization constraints on the worldsheet.  Thus, when the shell of branes sits at large radius, the non-abelian excitations are extremely heavy (regardless of the value of $k$) and can be integrated out; the assumed abelianized form of the brane effective action is valid.%
\footnote{The density of branes on the five-sphere at a given large radius is low, so even if the branes in the shell should encounter one another, that encounter is perturbative.  It is only when the shell is compressed to high density that strong coupling effects kick in.}
The W-strings become light when the shell approaches the scale at which an apparent horizon is forming.  It is well-known that the deconfinement transition in the gauge theory is dual to the Hawking-Page transition in gravity~\rcite{Witten:1998zw}; we are simply engineering that transition by moving the brane dynamics onto the Coulomb branch of the gauge theory where the non-abelian dynamics is made massive and decouples, and then collapsing the branes toward the origin of the Coulomb branch and seeing the transition happen, using a bulk description which is valid when the gauge theory is strongly coupled.

Note that we are not claiming that the {\it entire} gauge theory dual of bulk gravity is abelian~-- rather, the claim is that there is an {\it effective action} description of {\it particular regimes} in the configuration space which are well-described by a sum of the effective actions for isolated D3-branes coupled to bulk gravity.  When these isolated D3-branes, as well as bulk gravity in their vicinity, are not excited to the energy scale of the W-strings, those non-abelian excitations are so heavy that they are forced to sit in their ground state, leaving just the collective field theory of singlet excitations, which consist of bulk supergravity, coupled to the collective motions of the D3-branes themselves, which are governed by the $N$ copies of the DBI field theory as in~\eqref{jointeffact}.%
\footnote{This point ought not to be controversial.  We can for instance consider $g_s=10^{-5}$, with $N_0=10^{25}$ branes, separated by Coulomb vevs say of order $10^{30}$, apart from a particular cluster of $N=10^{10}$ branes which is excited to some temperature much smaller than these vevs.  While that cluster is strongly coupled, the remaining well-separated branes are not, and the parts of the scalar field space at scales much larger than the vev is completely unpopulated by excitations, with all these UV degrees of freedom sitting in their ground state.  The fact that there is an $AdS_5\times\bS^5$ geometrical description of this UV vacuum seems quite irrelevant (so to speak).  The whole point of the Wilsonian approach is that we can integrate out these degrees of freedom, and they play no role in the effective field theory that governs low-energy dynamics.}

The branes couple to the ambient spacetime geometry through their energy-momentum tensor
\be
T^{\mu\nu}(x) = -\frac{\tau'_p}{\sqrt{-G(x)}}  \int \! d^{p\tight+1}\!\xi \,e^{-\Phi}\sqrt{-\gamma}\, \gamma^{ab}\partial_a X^\mu\partial_b X^\nu  \,\delta^{(6)}\big(x-X(\xi)\big)
\ee
In particular, the energy density $T^{00}$ of the branes involves $\dot T$; after substituting Eq.~\eqref{vels}, one finds
\begin{align}
T_{00} 
&= \frac{ \tau_p e^{-\Phi} V_{\sst\sfM}}{\Lambda^{1/2}} \int\! d\xi^0 \, \Big(\frac{\dot R^2}{f(R)}+2\frac{\Lambda}{\ell^2} \Big)\delta^{(6)}\big(x-X(\xi)\big)
\end{align}
while the spatial components are
\be
T_{ij} = \gamma_{ij}\bigg[\tau_p e^{-\Phi} V_{\sst\sfM}\Lambda^{1/2} \int\!d\xi^0\,\delta^{(6)}\big(x-X(\xi)\big) \bigg] ~.
\ee
We can again absorb $\Lambda$ into the scale of the worldvolume coordinates.

The D-brane is a point source in the six-dimensional transverse space.  In order to avoid the complications of precisely where they are on the angular $\bS^5$, we simplify matters by averaging their locations over the sphere, assuming a roughly uniform shell of branes.  This averaging of the harmonic function over $\bS^5$ produces an effective domain wall source in $AdS_5$, and in the process introduces a numerical factor of $2/3$ from the averaging.%
\footnote{A similar averaging was used in the analysis of~\rcite{Brown:2024ajk}.}

The dynamics of such domain walls has been considered in a variety of contexts; a partial list of references includes~\rcite{Chamblin:1999ya,Kraus:1999it,Stoica:2000ws,Mukohyama:2000ga,Kofinas:2001uq,Barbon:2010us}.  We will in particular follow the treatment of~\rcite{Stoica:2000ws}, which allows for the possibility that both the effective 5d Newton constant and the 5d cosmological constant jump across the domain wall, as they do in the present context.


In the Einstein equations, the spacetime metric is continuous at a domain wall, but its gradient is discontinuous.  This discontinuity is captured by the Israel junction conditions~\rcite{Israel:1966rt}, which in our case read (taking into account the numerical factor of 2/3 from source averaging, and the jump in Newton's constant and the curvature radius across the domain wall)
\be
\label{israel}
\frac23\bigg(\frac{K_{ij}^+}{G_5^+} - \frac{K_{ij}^-}{G_5^-}\bigg) = -8\pi\Big( T_{ij}-\frac13 T\gamma_{ij}\Big)
= -\frac{8\pi}{3}\tau_3 \gamma_{ij} 
~.
\ee
Here $K_{ab}^\pm$ are the extrinsic curvatures of the brane as measured from either side, determined by the Lie derivative of the spacetime metric along the normal to the brane
\be
\label{Kdef}
K_{ab} = \half\partial_a X^\mu\partial_b X^\nu\cL_n G^*_{\mu\nu}~.
\ee
The factor of $2/3$ on the LHS of~\eqref{israel} comes from the aforementioned 
$\bS^5$ average.

The tangent $u^\mu$ and normal $n^\mu$ to the brane in the $r$-$t$ plane are given by
\begin{align}
\label{undef}
u^\mu &= \frac{1}{\ell\,(f\dot T^2\tight-\dot R^2/f)^{1/2}}\Big(\dot T,\dot R,\vec 0\Big)
\nn\\[.2cm]
n^\mu &= \frac{-1}{\ell\,(f\dot T^2\tight-\dot R^2/f)^{1/2}}\Big(\frac{\dot R}{f},f\,\dot T,\vec 0\Big)
\end{align}
The spatial components are particularly simple in the present situation:
\begin{align}
\label{Kspatial}
K_{ij} &= \gamma_{ij}\Big(n^\mu\partial_\mu \log(R)\Big)
\nn\\[.2cm]
n^\mu\partial_\mu \log(R)&= \frac{-1}{\ell\,R\big(f(R)\dot T^2-\dot R^2/f(R)\big)^{1/2}}\Big( \frac{\dot R}{f(R)}\partial_t +f(R)\dot T\partial_r\Big) R
\\[.2cm]
&= -\frac{1}{\ell\,R} \sqrt{f(R)+(\partial_\tau R)^2}
\nn
\end{align}
where $\tau$ is the proper time of the branes' motion, and we have used the constraint~\eqref{hamcon} as well as the relation $\sqrt\Lambda\,d\xi_0=d\tau$ in the gauge~\eqref{tempgauge}.

The spacetime~\eqref{D3geom} is sourced by a collapsing shell of D3-branes.  There are several parameters in the initial state.  First, the values of $g_s$ and the number $N$ of branes in the shell (\ie\ the coupling and rank of the dual gauge theory) and their energy.  
There are corresponding parameters in the bulk dual -- $g_s$ and $N$ determine the asymptotic $AdS_5\times \bS^5$ radius of curvature $\ell$ and the Newton constant.  Some useful relations are
\be
\label{params}
G_{10} = \frac{g_s^2\lstr^8}{32\pi^2} = G_5\Omega_5 = G_5\big(\pi^3\ell^5\big)
~~,~~~~
\ell^4 = \frac{Ng_s\lstr^4}{4\pi^3}
~~,~~~~
\tau_3 = \frac{2\pi}{g_s\lstr^4}
~.
\ee
In particular one has
\be
\frac{1}{\ell_\pm G_5^\pm} = N_\pm\,4\pi \tau_3~.
\ee

Working in proper time $\tau$, the junction condition~\eqref{israel} becomes
\be
\label{is2}
N_+\bigg[f_+(R)+\Big(\frac{dR}{d\tau}\Big)^{\!2}\bigg]^{\half}
-N_-\bigg[f_-(R)+\Big(\frac{dR}{d\tau}\Big)^{\!2}\bigg]^{\half}
= \big(N_+\tight-N_-\big) R ~.
\ee
Substituting $N_\pm = \bar N\pm dN/2$, $\mu_\pm=\bar\mu\pm d\mu/2$ and solving for the velocity of the shell yields, at leading order in $N$
\be
\Big(\frac{dR}{d\tau}\Big)^2 = 
 -k + \frac{1}{R^2}\Big(\bar\mu+\half\bar N\frac{d\mu}{dN}\Big)
 +\half\Big(\Big[2\bar N \frac{d\mu}{dN} + R^4\Big]^{\half}-R^2\Big)
~;
\ee
expanding at large $R$,
\be
\Big(\frac{dR}{d\tau}\Big)^2 \sim
-k + \frac{1}{R^2} \Big(\bar\mu+\bar N\frac{d\mu}{dN}\Big) +\frac{1}{R^6}\Big(\frac{\bar N}{2}\frac{d\mu}{dN}\Big)^2+\dots
\ee
Comparing to~\eqref{vels} (remembering that worldsheet time is proper time after rescaling to absorb $\Lambda$), we identify
\be
\frac{2\Pt}{m} = \bar N \frac{d\mu}{dN} ~.
\ee
In principle, one could integrate this result over a finite thickness shell.  Instead, we will make the further approximation that all the branes are in a single thin shell, to simplify the analysis.

\subsection{Bounces vs. black holes}
\label{sec:bounce}

Consider the motion of a shell of branes, sprinkled uniformly over the five-sphere, and having a radial spread $\delta r$, as in figure~\ref{fig:CappedThroat}.  Outside the shell, one has the vacuum geometry~\eqref{D3geom}.  A Gaussian sphere at large radius surrounds all the fivebranes, and measures $N$ units of magnetic five-form flux, which supports the size $\ell$ of $\bS^5$.  As anticipated above, upon entering and passing deeper into the shell, the Gaussian sphere surrounds less and less brane charge, and so the radius of $\bS^5$ begins to shrink; at the inner edge of the shell, the $\bS^5$ shrinks away completely, and the geometry caps off.  In the junction conditions for smeared brane shells, this shrinking of the angular spheres is encoded in the change in the 5d Newton constant~\eqref{params} as one crosses the shell.

If we make a thin shell approximation for the entire distribution of infalling D3-branes, we can set $N_-=0$ in~\eqref{is2}.  Then one has
\be
\label{mu and v}
\mu = R^2\Big(k + \big(\partial_\tau R\big)^2 \Big)  ~.
\ee
The branes' mass increases with radius, so for a given radial momentum they slow down (as they do for $k=0$), unless there is a driving force, as there is for $k=-1$.
For $k=0$, the mass parameter $\mu$ is just the kinetic energy of the shell as it moves along the flat potential; for $k=-1$, it is the sum of the kinetic energy plus the potential energy $-R^2$ of the branes.

\begin{figure}[ht]
\centering
\includegraphics[width=.7\textwidth]{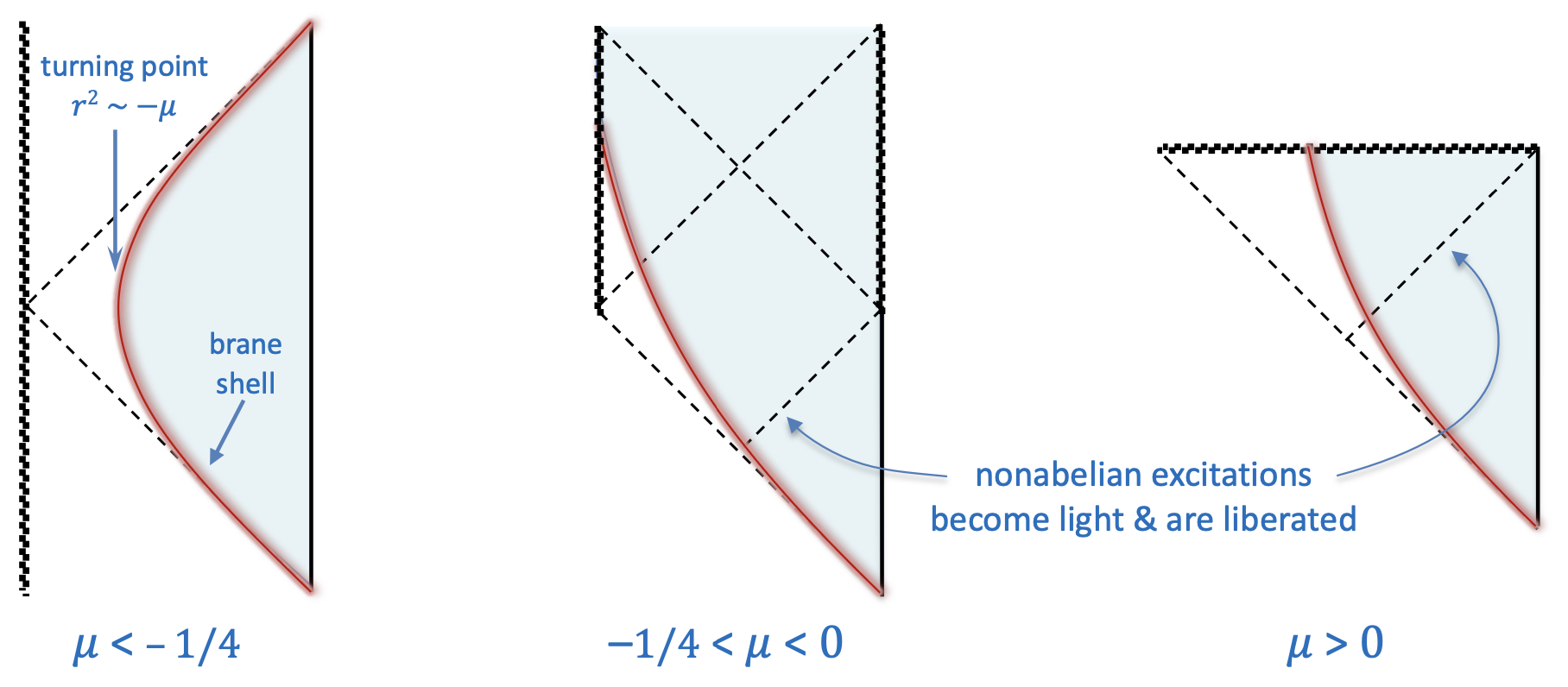}
\caption{\it Classical Penrose diagrams for the collapse of a hyperbolic D3-brane shell in various mass ranges.  Note that when $\mu<-1/4$, the singularity of the vacuum geometry at $r=0$ is never reached, as the geometry is always capped off by the shell at some much larger radius.}
\label{fig:hyperhistories}
\end{figure}

For $k=-1$, there is a turning point of the motion for sufficiently negative $\mu$ at 
\be
\label{turnpt}
R_*^2 \approx -\mu ~.
\ee
Such hyperbolic geometries cannot be stationary~-- the branes that form them are either infalling or outgoing.  
The shell will reach a minimum radius and then bounce back out, as depicted on the left in figure~\ref{fig:hyperhistories}.
%
%
%
%

Above the threshold $\mu=-1/4$, the collapse of the shell of hyperbolic D3-branes forms a black hole; similarly for positive mass toroidal D3-branes.  Approaching this threshold from below, we see that the turning point~\eqref{turnpt} doesn't quite match the extremal horizon radius $r_\pm^2=1/2$, perhaps due to the various approximations that have been made.

\subsection{Trapped shells and horizon formation}
\label{sec:horizon}

When $\mu$ lies above threshold, $\mu=0$ for toroidally wrapped D3-branes and $\mu=-1/4$ for the hyperbolic case, the brane shell reaches sufficiently small radius that the non-abelian excitations between the D3-branes become light and deconfine.  A trapped surface appears in the bulk geometry, and the branes are trapped inside a black hole.  Comparing~\eqref{mu and v} to the horizon scale of the black hole set by $f(r_+)=0$, we see that the proper velocity of the branes at the scale of horizon formation is
\be
\label{vel at hor}
(\partial_\tau R)^2 = r_+^2 ~.
\ee
Once the brane dynamics equilibrates, one has the black hole thermodynamics~\rcite{Emparan:1999gf}
\begin{align}
\label{BHthermo}
\ell M_\bh &= \frac{3N^2V_{\sst\sfM}}{8\pi^2 }\Big( r_+^4+kr_+^2+\frac{k^2}{4}\Big)
~~,~~~~
S_\bh = \frac{N^2 V_{\sst\sfM}}{2\pi}\, r_+^3
\\[.2cm]
\ell T_\bh &= \frac{r_+}\pi +\frac{k}{2\pi r_+}
\qquad\qquad\qquad~~,~~~~~\,
r_+ = \frac{\pi \ell T_\bh}{2}\bigg[ 1+\sqrt{1-\frac{2k}{\pi\ell^2T_\bh^2}}\ \bigg] ~.
\nn
\end{align}
Thus the shell's proper velocity is of order the scale of the Hawking temperature at the point of horizon formation.

The figure of merit for the likelihood of string creation in the collapsing shell is the rate of change of the string's energy relative to that energy
\be
\label{FOM}
\frac{\partial_t (\ell E_\str)}{(\ell E_\str)^2} ~.
\ee
The string energy $E_\str$ given in~\eqref{Estr} is proportional to $\sqrt{f(R)}$, thus such excitations cost very little energy when a trapped surface is forming, as the redshift factor $f(R)$ goes to zero.  On the other hand, from~\eqref{vels} 
\be
\partial_t (\ell E_\str)\Big|_{r\sim r_+} 
\propto \frac{f'(R)\,\partial_t R}{\!\sqrt{f(R)}} \approx 4\pi\ell T_\bh \sqrt{f(R)} ~,
\ee 
so~\eqref{FOM} indeed becomes large as the black hole starts to form.%
\footnote{One could alternatively evaluate this figure of merit using the Killing time $T$ rather than the proper time $\tau$ of the shell; in that case, one multiplies~\eqref{FOM} by $1/\dot T$ from~\eqref{vels}.  The production of strings is still growing as the shell approaches its horizon. }
On the other hand, in the early stages of collapse, at large radius we have $\sqrt{f(R)}\sim R$, and $\partial_t R\sim R$ from~\eqref{vels}, hence the figure of merit is vanishing like $1/R$ at large $R$.

Note that the shell may have some statistical distribution of nearest-neighbor brane separations.  If so, then the strings stretching between them will have some hierarchy of throat depths at which they become activated, and so small black holes may form somewhat further up the throat which then undergo a sequence of mergers, until the end state of a single large black hole is formed.%
\footnote{Such mergers were considered in a BPS context in~\rcite{Bena:2006kb}.}

Note that the W-string excitations whose energies are much larger than the temperature scale remain in their ground state, and can still be integrated out.  This corresponds to the fact that bulk gravity in the vacuum black hole geometry continues to describe physics well outside the thermal atmosphere of the black hole.  The non-abelian degrees of freedom that comprise the black hole state space are confined to the vicinity of the black hole.  The minimal set of degrees of freedom to capture the black hole entropy and the immediate vicinity of the horizon are the order $S$ non-abelian modes at the scale of the temperature and below, coupled to bulk gravity.  The latter is the collective field theory of the singlet excitations above the scale of the temperature.

The black D3-brane geometry~\eqref{D3geom} describes the long-lived intermediate state resulting from the collapse of the brane shell.  This state cannot decay via the usual emission of supergravity Hawking quanta, which are bound to the black hole in a thermal atmosphere due to the confining nature of the $AdS_5\times \bS^5$ geometry for such excitations.  

The underlying branes, however, came together along their Coulomb branch, and so they can go out again along their Coulomb branch.  The Hawking radiation of such branes is however highly suppressed thermodynamically~-- exponentially in $N$, in fact, as we will discuss presently.  The upshot is that the black hole is a long lived metastable brane bound state that decays very slowly by brane emission.  The main distinction between the toroidal and hyperbolic brane topologies is that the Coulomb branch has a flat potential in the toroidal case, so that the emitted branes only slowly move away from the black hole as a kind of ``holar wind'', while in the hyperbolic case the Coulomb branch is unstable, so that as soon as an emitted brane gets sufficiently far from the black hole it is rapidly swept away toward the conformal boundary of spacetime.


\section{Hawking radiation of D-branes}
\label{sec:hawking}

The heuristic picture of Hawking radiation is that it proceeds by a vacuum fluctuation near the horizon nucleating a particle-antiparticle pair, with the pair separating in the near-horizon gravitational field to the point that the particle emerges from the black hole and travels away toward spatial infinity, while the antiparticle member of the pair remains behind, just inside the horizon in a negative energy state.

\begin{figure}[ht]
\centering
  \begin{subfigure}[b]{0.35\textwidth}
  \hskip 0cm
    \includegraphics[width=\textwidth]{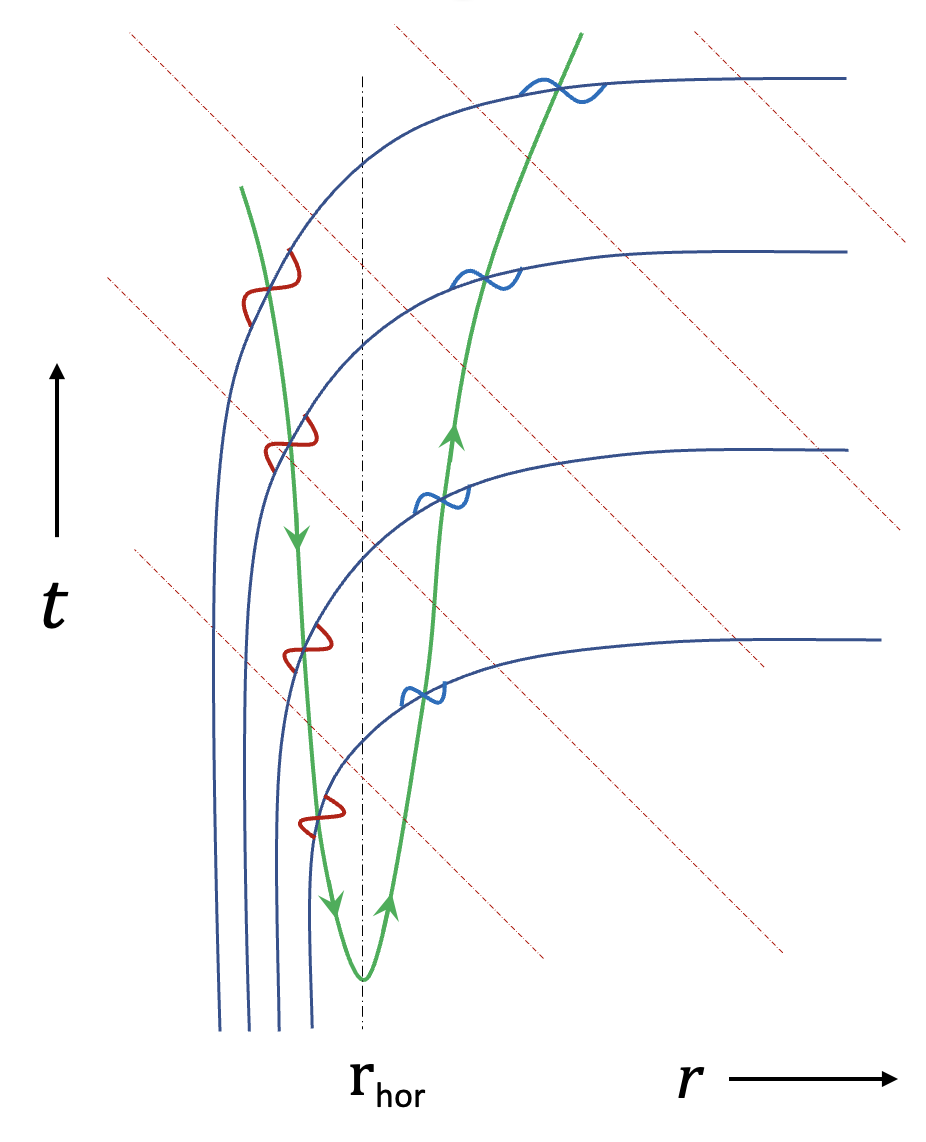}
    \caption{ }
    \label{fig:HawkingPtcl}
  \end{subfigure}
\qquad\qquad
  \begin{subfigure}[b]{0.35\textwidth}
      \hskip 0cm
    \includegraphics[width=\textwidth]{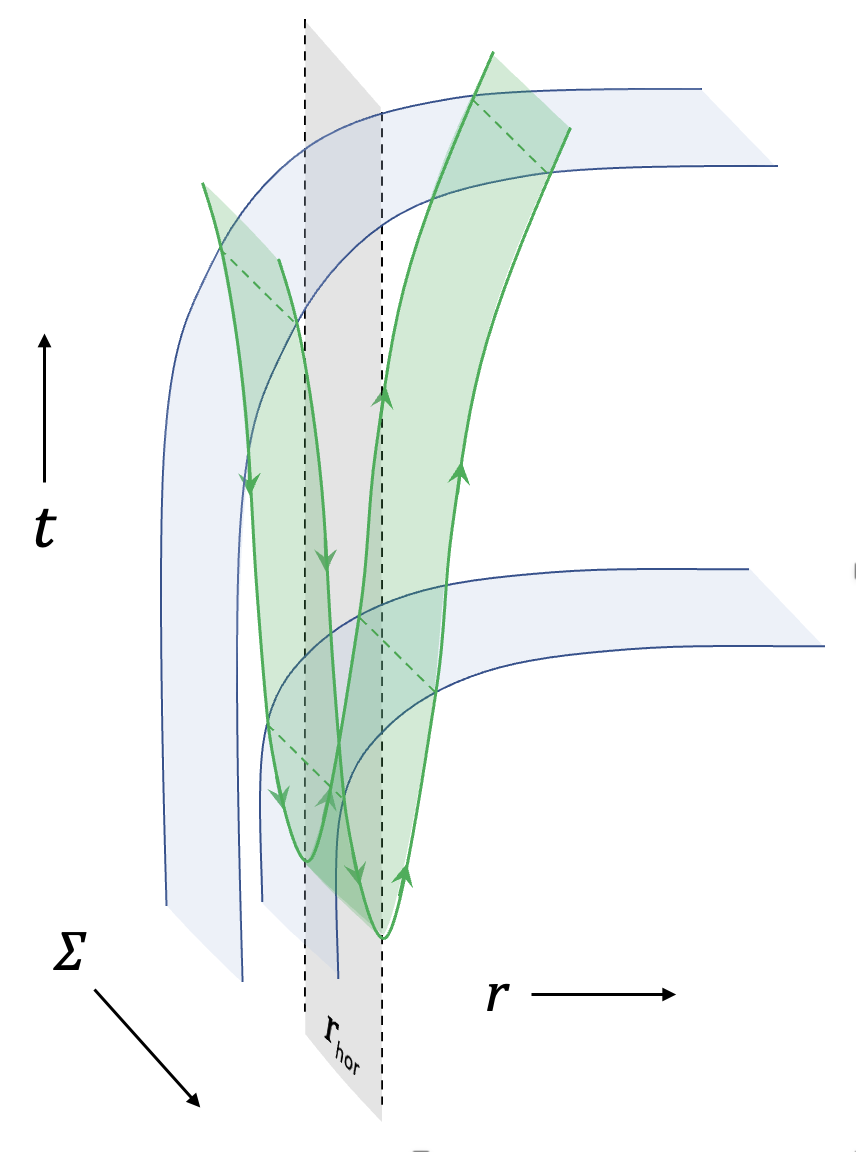}
    \caption{ }
    \label{fig:HawkingBrane}
  \end{subfigure}
\caption{\it 
Hawking radiation according to Hawking:
a) A vacuum fluctuation near the horizon (depicted here in coordinates that smoothly slice through the horizon) separates in the near-horizon gravitational field and becomes a particle-antiparticle pair, with the particle member of the pair escaping.
b) Brane Hawking radiation results from the analogous process of brane-antibrane pair creation near the horizon.  The branes are shaded green; the gray-shaded region is the horizon, and spacelike hypersurfaces are shaded blue.  The front and back of the diagram are periodically identified. 
}
\label{fig:HawkRad}
\end{figure}

In the tunneling approach pioneered in~\rcite{Kraus:1994by,Kraus:1994fj,Parikh:1999mf}, one assumes that a particle or brane materializes just inside the horizon, and computes the probability for the particle/brane to tunnel across the horizon so that it can escape as a Hawking quantum.  The tunneling D3-brane is again treated as a charged shell of energy $E$.  
As such, it starts its life inside a black hole of mass $M$ and D3-brane charge $N$ and leaves behind a black hole of mass $M-E $ and one unit less of D3-brane charge, and so the tunneling portion of its worldline starts just inside the initial black hole horizon and ends just outside the horizon of the final black hole.  One can further follow the evolution outside to compute the graybody factor in a WKB approximation.

A useful feature of this approach is that it is agnostic about where the tunneling brane came from, and so that analysis can be thought of either as pertaining to the standard Hawking process, where it arises from vacuum pair creation; or as an effective field theory description of a brane emerging from horizon-scale microstructure.

\subsection{Emission amplitude}
\label{sec:tunnelprob}

The calculation is typically done in Painlev\'e-Gullstrand coordinates, whose AdS version is described \eg\ in~\rcite{Hemming:2000as,Wu:2006pz}, though any non-singular coordinate system (for instance Eddington-Finkelstein) will do.  Starting from the coordinates $t,r$ of~\eqref{D3geom}, one reparametrizes the time coordinate to remove the coordinate singularity at the outer horizon via the transformation
\begin{align}
\label{paincoords}
dt = d\taub- \frac{\sqrt{1-F(r)}}{f(r)}\, dr
\end{align}
where one takes%
\footnote{There is freedom in the choice of $f_0(r)$; for instance, reference~\rcite{Hemming:2000as} makes the choice $f_0=1+r^2$ related to global AdS, while~\rcite{Wu:2006pz} included also a $1/r^2$ term.  The choice doesn't affect the near-horizon properties of the metric which determine the tunneling rate.}
\be
F(r)=\frac{f(r)}{f_0(r)}
~~,~~~~
f_0(r)=r^2 
~.
\ee
The resulting $r$-$\taub$ plane metric is
\begin{align}
\begin{split}
\label{Paincoords}
ds^2 &= \ell^2\Big[ -f\,d\taub^2 + 2\sqrt{1-f/f_0}\, d\taub\,dr + \frac{dr^2}{f_0} 
\,\Big]  ~.
\end{split}
\end{align}

The tunneling amplitude is given by the imaginary part of an effective particle action obtained by reducing on the spatial worldvolume of the brane 
\be
\label{Sptcl}
\cS_{\ptcl} = \frac m2 \int \!d\xi \,R(\xi)^3 \sqrt{\gamma}\Big[ \gamma^{-1} G_{\mu\nu}(x) \partial_\xi x^\mu \partial_\xi x^\nu -  1 \Big]
-q \int \! A_\mu(x) dx^\mu ~,
\ee
where $x^\mu=(T,R)$, $q=m$ and $A_\mu$ is the R-R gauge field reduced on $\sfM_3$, and $R^3V_{\sst\sfM}$ is the determinant of the spatial metric on the brane worldvolume.
In the Hamiltonian path integral, one has
\be
\cS_\ptcl = \int \!d\xi\, \Big( p_\mu \dot x^\mu - \lapse\, \sfH(p,x) \Big)
\ee
where the Lagrange multiplier $\lapse=\sqrt{\gamma}/2$ enforces the Hamiltonian constraint~\eqref{hamcon}.
We consider the tunneling portion of the trajectory near the horizon, which gives rise to a tunneling amplitude which is the imaginary part of the action.  
Solving the Hamiltonian constraint leaves the reduced action
\begin{align}
\begin{split}
\label{ImS}
\Im \cS_\ptcl &= \Im \Big[ \int_{r_{\rm in}}^{r_{\rm out}} \! dr \, p_r  \Big] 
\end{split}
\end{align}

Near the horizon, the radial momentum of the outgoing trajectory has a pole
\be
\label{prptcl}
p_r = \frac{r_+ ( \Pt + \mhat\, r_+^4) }{(r_+^2-r_-^2)(r-r_+)}  ~+~ {\it nonsingular}
\ee
where $m$ is defined in~\eqref{mdef}.
This pole is not a coordinate artifact, since we are working in non-singular coordinates near the horizon.  Rather, it is tied directly to the fact that the horizon is the surface of infinite redshift, where $g^{~}_{\sst\cT\cT}$ vanishes.  Indeed, the above expression comes from the Taylor expansion of $g^{~}_{\sst\cT\cT}$ near the horizon, in which the coefficient of the leading term (and thus the residue of the pole) involves the surface gravity
\be
\kappa = -\half \big[\partial_r g^{~}_{tt}\big]_{r=r_+} = \frac{r_+^2-r_-^2}{r_+} ~.
\ee

The pole along the integration contour is resolved by deforming it slightly into the lower half plane
and using the principle value prescription $\frac{1}{r-i\eps} = \cP( \frac1r) +i\pi\delta(r)$. 
The imaginary part of the integrand is thus
\be
\label{DeltaS}
\frac{\pi r_+}{r_+^2-r_-^2} \big(\Pt  + \mhat r_+^4 \big) = - \frac{dM
+  q^{\vphantom{|}}_{D3} C^*(r_+) }{2T} = -\half\,dS_\BH
\ee
where we have used conservation of energy and D3-brane charge in spacetime, as well as the first law of black hole thermodynamics and the relation between black hole surface gravity and temperature
\be
\kappa = 2\pi \ell T_\bh ~.
\ee
One finds a remarkably simple and universal answer: The imaginary part of the action is half the total change in the black hole entropy during the emission process.

We can rewrite~\eqref{DeltaS} as 
\be
\label{DeltaSalt}
\frac{\pi r_+}{r_+^2-r_-^2} \big(\Pt  + \mhat r_+^4 \big)
= \frac{\pi r_+}{r_+^2-r_-^2} \delta E + \pi m r_+^3  
\ee
where
\be
\label{E0}
\delta E = \Pt-E_0
~~,~~~~
E_0 = - mr_+^2r_-^2 = m \mu \propto \frac{M}{N} ~, 
\ee
and we have used~\eqref{horrad}.
This rewrites $dS_\BH$ in~\eqref{DeltaS} in terms of the thermal suppression of brane energy $\delta E = E-E_0$ above the reference value $E_0$, together with a term $2\pi mr_+^3$ which is the cost of producing a D3-brane wrapping the horizon.%
\footnote{This rewriting comports with the way things work in the BTZ case worked out in~\rcite{Martinec:2023plo}.  In that analysis, $E_0$ turns out to be the ``spectral flow energy'' of extracting a string out of the flux background, and $2\pi n_5r_+$ is the cost of producing a unit winding fundamental string at the horizon. }

It is worth pausing here to remark that this result is a perfectly natural consequence of basic principles of quantum theory and thermodynamics.  Fermi's Golden Rule states that the transition probability under the action of a generic operator $\cO$ (for instance, the Hamiltonian) is given~by
\be
\label{FGR}
\Gamma = \sum_{f}  \Big| \big\langle \psi_f\big|\cO \big|\psi_i\big\rangle \Big|^2 ~, 
\ee
where the transition matrix elements between an initial black hole state $\ket{\psi_i}$ and a final state $\ket{\psi_f}$, consisting of a slightly smaller black hole plus an emitted Hawking quantum, should be governed by the Eigenstate Thermalization Hypothesis (ETH)
\be
\label{ETH}
\big\langle \psi_f\big|\cO \big|\psi_i\big\rangle \sim \cF(\cO)\,\delta_{fi} + e^{-S_i/2}\,\cR_{fi}
\ee
where $\cF(\cO)$ is some smooth function on the state space, $\cR$ is a random matrix, and $S_i$ is the entropy of the initial state.  Plugging this ansatz into the transition amplitude~\eqref{FGR} and using the fact that the magnitudes of the transition matrix elements are roughly uniform over the space of final states yields the result~\eqref{ImS}, \eqref{DeltaS}.
That we should choose the entropy of the initial state rather than some other value in~\eqref{ETH} is dictated by the highly absorptive nature of black holes.  If we consider the reverse process in which a slightly smaller black hole with a quantum outside makes a transition to a larger black hole by absorbing the quantum, the (adjoint) matrix element is now approximated by ${\it exp}[-S_f/2]$, which when squared cancels against the density of final states leading to an absorption probability of order one.  In the calculation sketched above, this reverse process corresponds to choosing the other branch of the solution for the radial momentum resulting from the Hamiltonian constraint, which (a) corresponds to the ingoing trajectory, and (b) has no pole at the horizon and therefore no imaginary part.

While we have not incorporated the back-reaction of the emitted quantum on the ambient near-horizon geometry, it is possible to solve the constraints of the bulk gravity theory coupled to the worldvolume dynamics of the escaping quantum~\rcite{Kraus:1994by,Kraus:1994fj,Keski-Vakkuri:1996wom,Parikh:1999mf,Massar:1999wg}.  The constraints can be boiled down to a coupling between the radial dynamics of the quantum in terms of the conjugate pair $r,p_r$ on the worldline, and the conjugate pair of gravitational variables given by the horizon area $A$ and the Rindler-like time $\Theta$ near the horizon.  This coupling relates the change in the horizon area to the worldline reduced action beyond the linearized level, and extends the result~\eqref{Gambh} beyond small changes in the entropy~\rcite{Kraus:1994by,Kraus:1994fj}.   

Once emitted via the averaged transition probability $\Gamma\sim \exp[{\delta S_\BH}]$, a Hawking quantum encounters the ambient geometry.  There is a transmission amplitude for the quantum to escape to infinity, and a reflection amplitude to be re-absorbed by the black hole.  This is the graybody filter that surrounds any black hole (see \eg~\rcite{Harmark:2007jy}), whose effects are not incorporated in the above calculation, but could be included by a more detailed analysis of the path integral for the escaping quantum as it propagates away from the very near-horizon region.  For instance, supergravity quanta in the decoupling limit lie in bound state wavefunctions, and so will always reach some maximum radius characterized by their energy and fall back into the black hole.
On the other hand, charged D3-branes with sufficient initial radial momentum will escape onto the Coulomb branch of the D3-branes and travel to spatial infinity along a timelike trajectory.

The full expression for the radial momentum is
\begin{align}
\label{fullPr}
p_r &= \frac{\Big( \big(\Pt+\mhat r^4\big)\sqrt{r^2(r_+^2+r_-^2)-r_+^2r_-^2} 
+ r^2 \sqrt{(\Pt+\mhat r^4)^2-\mhat^2 r^4(r^2-r_+^2)(r^2-r_-^2)}
\,\Big)}{(r^2-r_+^2)(r^2-r_-^2)}
\nn\\[.2cm]
&= \frac{\Big( \big(\Pt +\mhat r^4\big)\sqrt{\mu-k r^2} 
+ r^2 \sqrt{\Pt^2+(2\Pt +\mu\mhat)\mhat r^4 -k \mhat^2r^6}
\,\Big)}{(r^2-r_+^2)(r^2-r_-^2)} ~.
\end{align}

Consider this quantity expressed in terms of the relative energy $\delta E=E-E_0$.  For the toroidal case $k=0$, the argument of the second square root in this expression can be written (using the fact that $\mu=r_+^4$ for $k=0$)
\be
E^2+(2E+mr_+^4)mr^4 = \delta E^2 + (2\delta E-mr_+^4)(r^4-r_+^4) ~,
\ee
which is negative at large $r$ unless $\delta E = E-E_0> \half m r_+^4 = \half m\mu\propto M/N$.  Branes with less than this threshold energy are forbidden from escaping the black hole.  Branes whose energy exceeds this threshold can escape, but this means that they carry away an energy of order $1/N$ of the remaining black hole's mass.  The fact that there is an energy above threshold for a brane to escape to infinity along the Coulomb branch reflects the fact that the system is well above the BPS bound~-- the gravitational attraction of the black hole exceeds by an order one amount the electrostatic repulsion between brane charges, and the branes must have sufficient kinetic energy to overcome this force imbalance in order to escape.

Note that any excess energy above threshold is heavily suppressed thermodynamically; it takes away additional free energy that further decreases the density of states available after emission.  This includes any energy that might be used to excite internal modes on the escaping brane, or to impart any substantial radial momentum that could expeditiously take the emitted brane away from the black hole.  Emitted branes are produced in their (256-fold degenerate) ground state, very near threshold. 

For $k=-1$, when the black hole mass is large, $\mu\gg1$, and the brane energy is much less than this scale, $\delta E\ll\mu$, one finds that the second square root in~\eqref{fullPr} is imaginary over the range
\be
\mu^{\frac14} \lesssim r \lesssim \mu^{\frac12}
\ee
with a barrier height
\be
\Delta V_\eff \sim 
\mu^{\frac32}\,m
\ee
that highly suppresses the escape of emitted branes in this energy range~-- the graybody transmission amplitude is essentially zero.  On the other hand, the barrier again completely disappears for $\delta E\gtrsim \half m\mu$.  Thus, at large $\mu$ the graybody potential essentially completely suppresses D3-brane emission, until one reaches into the tail of the Hawking emission spectrum at energies of order $m\mu$.

Note that this energy is of order $1/N$ of the total black hole mass.  As a result, it is possible to radiate an order one fraction of the initial black hole mass through the emission of the $N$ branes.  One might have worried that a modest graybody modification of the emission spectrum would result in the emitted branes being only slightly above the threshold for emitting a D3-brane, in which case the black hole would lose a large amount of entropy but very little mass; the remaining black hole would have been driven further and further from extremality with each brane emission, leaving almost all of the energy to be radiated in the final stages of the decay.  The estimates above indicate that an order one fraction of the mass is radiated before the end stage, but the precise amount will depend on the precise profile of the graybody reflection and transmission coefficients, which we will not attempt to pin down here.

As in the toroidal case, the radiated branes will be predominantly emitted very near the threshold where the brane just barely spills over the graybody potential.  So the brane starts off without internal excitations, almost at rest at the top of its effective potential, and then falls away to spatial infinity.  

Since the gauge-theoretic description of this escape of a brane onto the Coulomb branch consists of the de-excitation of all the non-abelian modes that bind it to the other branes in the black hole, one could worry that a few such strings might still dangle off the black hole as it escapes.  If there are only a few, they are incapable of overcoming the potential force that is pushing the brane out along the Coulomb branch, and so there might be a few remnant strings stretching between the escaping brane and the black hole left behind.  However, a calculation of the tunneling amplitude to produce a brane together with stretched strings~\rcite{Wenren:2017ryk} indicates that it is once again exponentially further suppressed beyond the amplitude for emitting a brane without such stretched strings.  Furthermore, the Gauss law on the emitted brane requires that an equal number of stretched strings and antistrings must be created, so it is always possible for them to annihilate with one another, leaving a departing brane with no strings attached (though perhaps slightly excited as a result of the annihilation process).

\subsection{Evolution history}
\label{sec:evolution}

We are now poised to describe the black hole S-matrix in examples of gauge/gravity duality endowed with a Coulomb branch of the gauge theory, along which the branes can escape the system.  

We begin with an infalling shell of branes in the asymptotic past.  At large $N$ and strong coupling, this shell is well-approximated by a sparse distribution of branes over the angular space of the scalars in the gauge theory, in some range of radii.  Outside the shell, the geometry is well-approximated by the vacuum geometry~\eqref{D3geom}.  The space transverse to the branes in this region has the structure of a cylinder $\bR_+\times\bS^5$.  As one dives into the shell, the angular sphere decreases in size and eventually pinches off smoothly at the inner edge of the shell, as depicted in figure~\ref{fig:CappedThroat}.

The shell is either moving in along a flat direction (for toroidally wrapped branes), or climbing an unstable effective potential (for branes wrapped on a compact hyperbolic manifold $\Sigma_3$).  On $\bT^3$, the collapsing shell always forms an apparent horizon and a black object, though at ultralow energies the horizon radius might be so small that the near-horizon region is better described in some other duality frame~\rcite{Martinec:1998ja}.  On $\Sigma_3$, if the shell has insufficient energy to climb all the way to the top of the effective potential, it reaches a turning point and falls back down, as depicted in the left-most diagram of figure~\ref{fig:hyperhistories}.  If it does have sufficient energy, again an apparent horizon forms, and the branes are trapped in a metastable black hole state at the top of the effective potential.  As discussed in~\rcite{Barbon:2010us}, if the energy above extremality is too low, the non-abelian brane excitations are ineffective at trapping the branes, and the metastable state decays back onto the Coulomb branch fairly rapidly.  If on the other hand the black hole mass is sufficiently large, the black hole lives for a time exponentially long in $N$.

The formation of an apparent horizon in the gravitational description is dual to a deconfinement transition in the gauge theory.  A given brane pair in the shell is separated in the angular space by $\delta\Omega_5$ and radially by an amount $\delta r$.  The energy of a string stretched between the two branes is of order $E_\str$, Eq.~\eqref{Estr} (plus a corresponding contribution from the radial separation of the branes).  This energy decreases with the redshift of the shell according to the lapse function $f(r)$, Eq.~\eqref{D3geom}.  For large mass $\mu\gg1$, one has from~\eqref{BHthermo}, \eqref{horrad}
\be
T_\bh \sim \frac{r_+}{\ell} \sim \frac{\mu^{\frac14}}{\ell} ~.
\ee
When the branes reach the scale of the apparent horizon, the kinetic energy of collapse~\eqref{vel at hor} thermalizes into stretched strings, and the gauge theory undergoes a deconfinement transition~\rcite{Witten:1998zw}.  

\begin{figure}[ht]
\centering
\includegraphics[width=.3\textwidth]{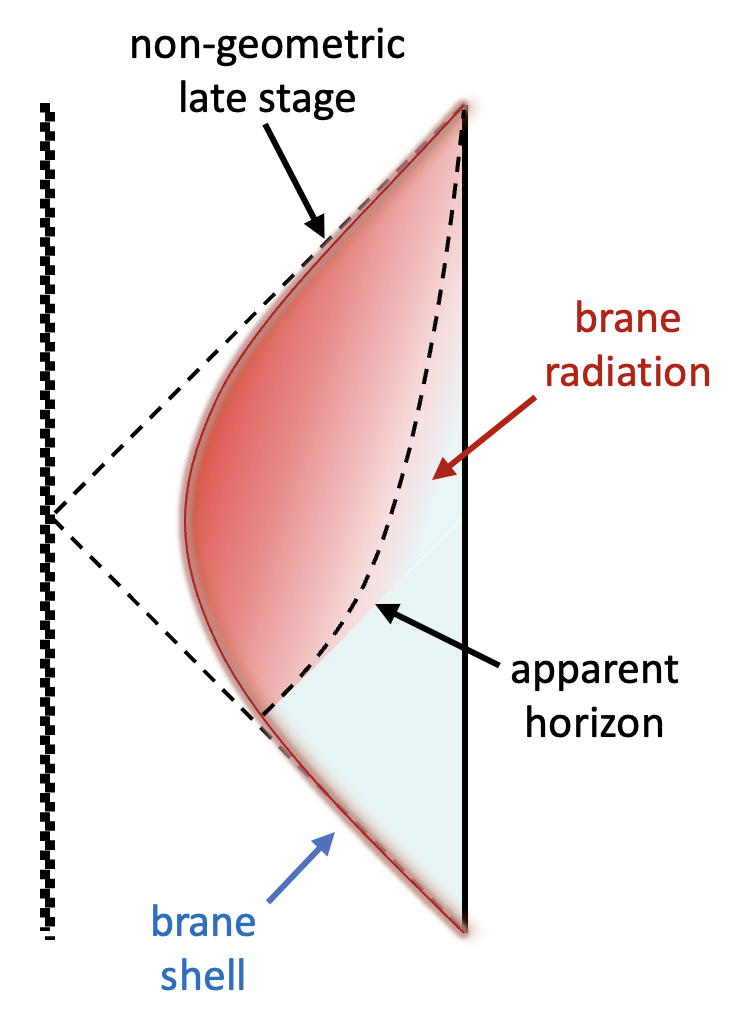}
\caption{\it Sketch of the Penrose diagram of the formation of a hyperbolic black hole from a collapsing brane shell, and its subsequent decay by re-radiation of the constituent branes.  The issue of what is happening in the interior lies beyond the scope of the effective field theory approach of the present work; see the discussion of section~\ref{sec:unitarity}, however.}
\label{fig:HyperPenrose}
\end{figure}

The black hole forms and quickly settles into equilibrium with its thermal atmosphere of supergravity quanta, which lie in $AdS_5$ bound state wavefunctions.  
The branes slowly radiate back onto the Coulomb branch via the tunneling process described in section~\ref{sec:hawking}.  Once radiated, they begin to travel toward the conformal boundary of spacetime.  A sketch of the associated Penrose diagram is given in figure~\ref{fig:HyperPenrose}.

%
%

For toroidally wrapped branes, the motion is along the flat brane potential; for hyperbolic branes, once they clear the near-horizon region they begin to accelerate away along the unstable brane potential.  Either way, at late times the individual branes are well out along the Coulomb branch of the gauge theory, which at any given time is spontaneously broken to $U(N)\times [U(1)]^{N_0-N}$, where $U(N)$ describes the black hole left behind, and the abelian factors describe the radiated branes.  As the evolution continues, the radiated branes are more and more decoupled from the black hole left behind, and so described again by a sum of $N_0-N$ DBI actions coupled to bulk supergravity, along the lines of~\eqref{jointeffact}.  
Exponentially rarely, another D3-brane manages to escape, and $N\to N-1$ in the effective bulk description.


\section{Unitarity}
\label{sec:unitarity}

The existence of gauge/gravity dualities with a Coulomb branch for the gauge theory allows us to consider the S-matrix in quantum gravity involving black holes as intermediate states, in situations where the existence of a field-theoretic dual description guarantees the unitarity of the S-matrix.  One can then contemplate what aspects of the bulk gravitational description are breaking down and need modification.  In particular, one may ask about where in the Hawking calculation of the out-state is a mistake being made.

\subsection{What is gravity missing?}
\label{sec:missing}

The bulk gravitational effective field theory simply doesn't have the necessary degrees of freedom to account for black hole entropy, and so at a minimum needs to be supplemented by those degrees of freedom in order to have any hope of tracking states from early to late times.  If you can't account for how information is stored, than there is no hope for accounting how it is recovered.

In the present context, these degrees of freedom are the $O(N^2)$ strings that stretch between the D3-branes in the black hole bound state.  We saw that, beginning with the abelianized brane effective field theory~\eqref{jointeffact} that describes the collapse phase, that effective field theory is beginning to break down when the collapsing brane shell is starting to fall through a trapped surface.  The trapping is associated to the liberation of strings stretching between the branes, which one can diagnose from their decreasing energy cost in the effective field theory as the brane shell compresses more and more, which in the back-reacted effective geometry is described as the shell descending a warped throat geometry. 

Gravity alone knows nothing about these degrees of freedom, and so necessarily it coarse-grains over black hole microstructure.
According to classical gravity, any ordinary matter near the horizon is inevitably swept into the black hole singularity, leaving a vacuum state near the horizon.  Any escaping brane then comes from a vacuum fluctuation near a smooth black hole horizon in which a brane-antibrane pair is formed.  The brane tunnels out, leaving an antibrane in a negative energy state just inside the horizon.

The WKB calculation of black hole radiance highlights the central role of the classical horizon; the entire emission probability%
\footnote{Apart from graybody factors, which can also be evaluated in the WKB approximation by continuing to follow the D-brane's evolution through the near-horizon region after it is emitted.}
comes from the infinitesimal neighborhood of the horizon, whose area determines the density of initial and final states, and thus the emission probability via~\eqref{Gambh}, \eqref{DeltaS}.

In the Hawking process, the branes being emitted are unrelated to those that initially collapsed to form the black hole.  Depending on how one chooses to slice the interior geometry, the initial branes either long ago fell into a singularity, or are sequestered on a ``nice slice'' deep in the black hole interior far from the horizon.  Either way, the emission of these branes would involve a tunneling amplitude that is much more highly suppressed than the Hawking process~-- they would have to traverse a spacelike trajectory to the vicinity of the horizon before tunneling across, so beyond the exponential suppression~\eqref{radrate1}, one would have an additional exponential (in $N$) suppression from this further non-classical path.  The Hawking branes are in a sense invented out of whole cloth, and as such convey no information about the initial state.

Mathur's small corrections theorems~\rcite{Mathur:2009hf,Guo:2021blh} then inevitably lead to a monotonically increasing Page curve in the Hawking paradigm, and a loss of S-matrix unitarity.  Each brane pair creation process increases the entanglement entropy between the radiated branes and the remaining black hole.

The details of the Page curve differ slightly from standard examples of black hole decay via Hawking emission.  Ordinarily, the number of emitted Hawking quanta is of order the entropy of the initial black hole.  For instance, the entropy of a 3+1d Schwarzschild black hole of mass $M$ is of order $(\ell_p M)^2$; the Hawking temperature is of order $M^{-1}$, and thus the number of quanta emitted by the time the black hole completely evaporates is also of order $(\ell_p M)^2$.  Therefore, the Page time occurs when about half of the Hawking quanta have been emitted.

But in gauge/gravity duality, the ordinary channel of emitting light supergravity quanta which then escape to a faraway region is closed off~-- supergravity excitations all lie in bound state wavefunctions that fall off radially on the scale of the $AdS$ radius.  
The only available channel for the out state is brane radiation, but that comes in huge chunks.  The initial entropy is of order $N_0^2$, and each emission reduces the entropy of the remaining charge $N$ black hole by an amount of order $N$ rather than order one.  And of course, since the radiation is extracting the $N_0$ initial $D$-branes one by one, there can be at most $N_0$ Hawking quanta.  The Page time occurs when the entropy of the remaining black hole and the entropy of the radiation are approximately equal, so that their entanglement entropy is the maximum it could possibly be under the assumption of unitary dynamics.  This occurs when $N^2\sim (N_0-N)$, or in other words when the remaining black hole consists of approximately $\sqrt{N_0}$ branes.

Nevertheless, there is still a Page time, and a conflict between the Page curve predicted by the Hawking process and that inferred from the unitary brane dynamics of the exact gauge theory dual.  
Supergravity modes constitute collective singlet excitations of the gauge theory.  While there is a thermal atmosphere of supergravity Hawking quanta, this atmosphere is subject to the same small corrections theorem, so if the modes in this atmosphere are generated by the Hawking process, they cannot convey a significant amount of information to the escaping branes.  And as the branes get further away from the black hole along their Coulomb branch, they are more and more decoupled from any excitations in the vicinity of the black hole.

Thus it would seem that the small corrections theorem will not allow two things to simultaneously hold:  First, that the Page curve comes down; and second, that the outgoing Hawking radiation is generated by the Hawking process of vacuum fluctuations in the vicinity of a smooth horizon that is well-described by supergravity alone, with matter (including string and brane matter) in its vacuum state at the horizon.

\subsection{How string theory differs}
\label{sec:differs}

String theory, on the other hand, describes the rare event of D-brane Hawking emission as a process in which all the strings that bind that particular brane to the $N-1$ other D-branes in the black hole simultaneously de-excite sufficiently that the brane is no longer bound, and can escape.  Bulk gravity cannot resolve this non-abelian microstructure, and so coarse-grains over it and describes the brane emission process instead as a vacuum fluctuation. 
String theory leads to a natural proposal for what non-vacuum structures near the horizon maintain unitarity.

In effective field theory, when degrees of freedom that have been integrated out become light enough that they enter the dynamics, one need not discard the effective theory; rather, one should simply add those degrees of freedom to the effective description and re-solve the effective dynamics.  In the present case, that means keeping the stretched strings at the scale of the black hole temperature and below.  One can still integrate out the stretched strings whose energies are at scales far above the black hole temperature, as these strings continue to sit in their ground state.  In other words, sufficiently far from the black hole thermal atmosphere, the bulk is accurately described by the vacuum black hole geometry.  By not integrating out the non-abelian strings below a certain scale, we have retained the string theoretic degrees of freedom which are responsible for the black hole entropy, and which maintain unitarity in the bulk description of the dynamics.
In gauge/gravity duality, it would seem that the minimal modification is simply to incorporate these light non-abelian degrees of freedom that are activated towards the end of the collapse epoch. 

In string theory, the brane being emitted {\it is} one of the original branes (highly scrambled due to the non-abelian dynamics of the intermediate black hole state).  The emission process does not {\it increase} from $N$ to $N+1$ the rank of the gauge theory describing the black hole after the emission of a brane, as in the Hawking description; rather, it involves a {\it reduction} in the gauge theory rank from $N$ to $N-1$.  Furthermore, the correct effective description includes the degrees of freedom responsible for the black hole entropy, so that one can track the evolution in the black hole state space from the initial infall to the subsequent radiation and decay.

The tunneling calculation can then be interpreted as a coarse-grained, thermodynamic approximation of an underlying exponentially rare microscopic process in which the non-abelian degrees of freedom associated to a given brane (and responsible for its contribution to the black hole entropy) all simultaneously de-excite, allowing the spontaneous breaking of the gauge group via the emission of a brane onto the Coulomb branch of the brane dynamics.

Note that these added degrees of freedom are {\it only} active on the scale of the black hole horizon; they are not acting non-locally between the black hole and the distant radiated branes.  The non-abelian modes that directly communicate between the black hole branes and the distant branes are so heavy, they are forced to sit in their ground state, and so cannot communicate any information, up to exponentially small corrections of the sort accounted for in the small corrections theorems.  

\begin{figure}[ht]
\centering
\includegraphics[width=.3\textwidth]{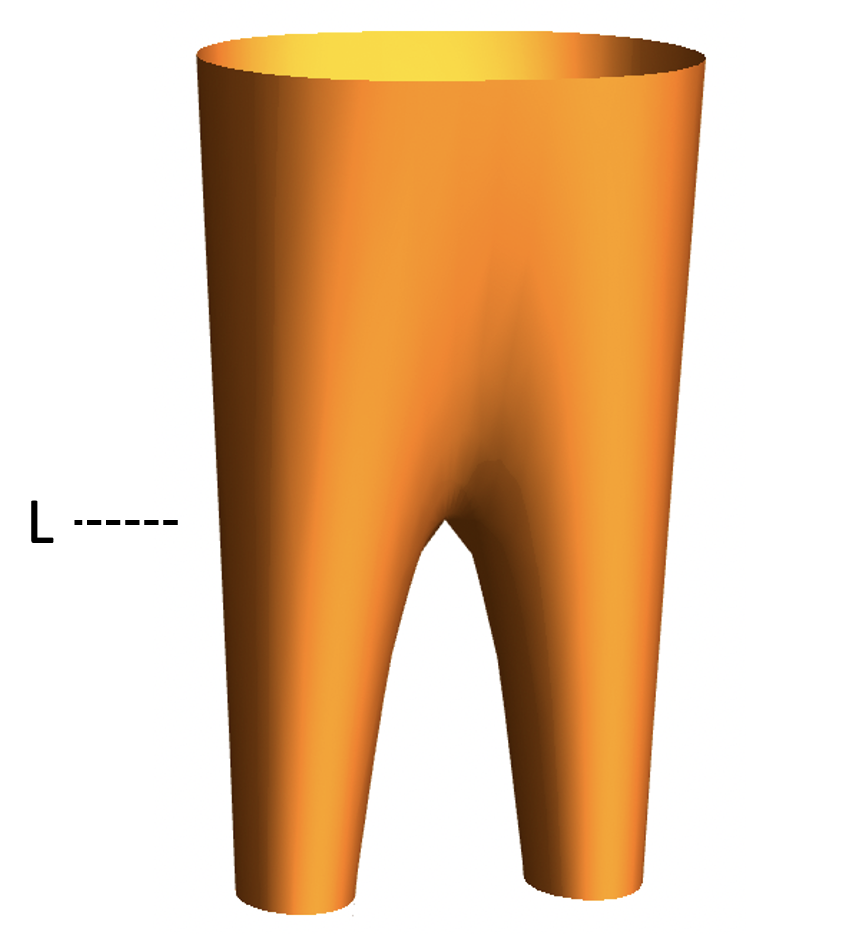}
\caption{\it A throat sourced by two clusters of D3-branes separated by a large Coulomb vev $L$ splits into two throats at a depth determined by $L$.  Communication between the two clusters is mediated by excitations traveling up one throat and down the other, which necessarily have energy of order $L$.  Tunneling effects are a tiny correction.  There are no non-local interactions, even at strong coupling.}
\label{fig:ThroatSplit}
\end{figure}

One might worry that there are some mysterious non-local effects that arise in the gauge theory at strong coupling,%
\footnote{Wormhole effects in the dual geometry constitute a popular example.}
however this idea flies in the face of our understanding of quantum field theory.  Consider $N$ toroidally wrapped D3-branes, separated into two clusters of $N/2$ branes well-separated on their Coulomb branch by some scale $L$.  The dual geometry inferred from~\eqref{HCoul} is asymptotically $AdS_5\times \bS^5$ with an angular (or curvature) scale set by $N$, which at the radial scale $L$ splits into two distinct $AdS_5\times\bS^5$ throats of angular size $N/2$, see figure~\ref{fig:ThroatSplit}.  The W-bosons that non-locally communicate between the two clusters of branes are forced to be in their ground state by the large scalar vev between the two clusters; the only way that observers at the bottom of the two throats can communicate is by sending a signal up their throat to where the other throat merges with it, and then back down that other throat.  But such a signal costs energy of order $L$, and so as $L$ is made larger and larger the two throats decouple, as expected.

%
%

Similarly, the degrees of freedom describing the black hole intermediate state in the Coulomb branch S-matrix have characteristic energy set by the temperature of the black hole, and so the branes radiated onto the Coulomb branch are decoupling from those degrees of freedom more and more as they run off to infinity.  This is particularly true of the hyperbolic black hole, where the radiated branes quickly accelerate to a constant velocity as they run off to the asymptotic region.  At any instant of time, one has a lopsided version of the above figure where one of the throats has the remaining black hole comprised of $N$ branes, and there are $N_0-N$ widely spaced single branes that have been radiated onto their Coulomb branch, which don't have their own throat geometries in their vicinity (their near-brane dynamics is on the stringy, non-geometric side of the correspondence transition~\rcite{Horowitz:1996nw}).  Nevertheless, the radiated branes decouple from the remaining black hole as they run away, for the same reason~-- it would take a Hawking quantum whose energy was of order the Coulomb vev to communicate with them.

It seems that given the validity of gauge/gravity duality, the horizon serves as a proxy in the effective theory for the unresolved black hole density of states, and is not a causal barrier per se in the theory at finite $N$ where the density of states is finite.  Classical black holes are black because an incident quantum is easily absorbed into a phase space whose volume diverges in the $N\to\infty$ limit, which it scrambles into.  The scrambled constituents then ergodically explore that phase space forever; the probability of return vanishes.  Quantum black holes at finite $N$ are not black, because the state space is finite dimensional, and the probability of return is therefore finite.  There is no causal barrier to the black hole emitting the information stored in it; the emission probability is the essentially kinematic factor~\eqref{Gambh} embodying the likelihood that energy and charge \etc.\ stored in the black hole leaks out into the much smaller region of phase space consisting of ordinary effective field theory quanta outside the black hole. 
The emission amplitude is a small tail near the horizon of the coherent wavefunction of the black hole's underlying entropic degrees of freedom.
As discussed in section~\ref{sec:tunnelprob}, the characteristics of the emission are governed by fundamental considerations of phase space, detailed balance, thermality, unitarity, \etc.

On the other hand, the Hawking calculation (1) doesn't resolve which final state one has, (2) averages over initial states, and (3) doesn't correctly account for the change in the entropic degrees of freedom in the microstates resulting from the transition.  It generates an incorrect entanglement structure between the emitted quantum and the remaining black hole, eventually replacing it by a density matrix.   

This issue is put into stark relief by the above calculation of the discharging of a D3-brane black hole by the emission of the underlying D3-branes.%
\footnote{Similar considerations apply to any situation in which Hawking radiation discharges a charged black hole.}
In the naive supergravity effective field theory, the $N$ infalling D3-branes in the background travel a trajectory that is well-separated from the antibrane Hawking partners clustered just inside the horizon.
The Gauss law for the antisymmetric tensor gauge theory under which the branes are charged keeps track of the initial infalling branes.  The singularity (or the infalling shell deep in the black hole interior, depending on the slicing) is the source of the tensor gauge flux after the initial collapse.
The Hawking process then draws brane-antibrane pairs from the vacuum near the horizon which are completely entangled with one another, and thus the emitted brane is not at all entangled or correlated with the original branes that formed the black hole, which are macroscopically spacelike separated from this Hawking process.  The emitted brane knows nothing about the microstate of the black hole it emerged from, according to this line of reasoning.

In the dual gauge theory, on the other hand, the emitted brane {\it is} one of the $N$ background branes in the initial bound state.  It is entangled with the remaining $N-1$ branes rather than with some antibrane vacuum fluctuation.  This suggests that the correct bulk description, rather than having vacuum structure at and just within the horizon, with the emitted branes arising from vacuum fluctuations near the horizon, instead has a wavefunction for the background branes with support at the horizon, and it is one of these branes that undergoes the tunneling process calculated above.   
If the branes emerging are the same branes that formed the black hole to begin with, then at any given moment some fraction of those branes has to be lurking near the horizon, waiting for their chance to escape their imprisonment.  And then the Gauss law detects their presence, and one does not have the supergravity effective field theory vacuum at and within the horizon.  Unitarity of charged Hawking radiation, and locality of the effective dynamics at the horizon scale, {\it requires} the charged matter emerging from the horizon to be the original branes, on-shell {\it at} the horizon, with a tiny probability governed by ${\it exp}[\delta S_\BH]$, Eq.~\eqref{DeltaS}.  
The ``tunneling'' amplitude calculated above represents some average over the microstate dynamics.
The emerging brane does not come from further inside a black hole with vacuum geometry in the interior, because then the brane would traverse a longer non-classical trajectory with a larger contribution to the imaginary part of the action~\eqref{ImS}, leading to a further suppression of the emission probability that is incompatible with the thermodynamics.%

Note that the direct description of the black hole interior lies beyond the scope of the effective field theory under consideration, which restricts its attention to the description of the black hole from the exterior perspective. Some past speculations on how string theory might describe the black hole interior are reviewed and discussed in Appendix~\ref{sec:interior}. 
The conclusion above, that the black hole interior is not approximately the vacuum, is an inference based on the small corrections theorem, the unitarity of the dynamics, and the Gauss law for the brane charge, rather than a direct calculation.

How can this be?  Ordinary matter is destined to collapse into a singularity, leaving vacuum geometry in its wake.  Perhaps the answer lies in the fact that the entropy in the black hole phase is associated to a deconfinement transition in the dual gauge theory~\rcite{Witten:1998zw}.  This suggests that the horizon is a phase boundary, and that rather than being in a quiescent, vacuum state, the black hole interior is in a non-geometric phase.  The gravitational degrees of freedom are a set of $O(N^0)$ collective modes of the gauge theory (associated to single-trace operators) that are swamped in the deconfined phase by the full $O(N^2)$ set of supergluons.  There are of order $N^2$ more species of excitation present than supergravity accounts for.  These excitations are both the source of black brane entropy, and serve to bind the branes together, making their escape exponentially unlikely.  But the plasma of internal brane excitations is expected to be in causal contact with the collective modes describing the black brane exterior, and influence the state of the thermal atmosphere so that Hawking quanta carry away entanglement with the branes that initially formed the black hole.  The emission of a Hawking quantum de-excites the existing excited interior degrees of freedom rather than exciting new degrees of freedom that were initially in a lower excitation state, maintaining consistency with the diminishing dimensionality of the black hole state space at each step.



\section{Final remarks}
\label{sec:discussion}

While our focus has been on the duality between $AdS_5\times \bS^5$ string theory and $\cN=4$ super Yang-Mills in 3+1d, similar considerations apply to other examples of AdS/CFT duality, such as the ABJM theory~\rcite{Aharony:2008ug} dual to \eg~type IIA string theory on $AdS_4\times\bC\bP^3$.  The calculations above go through essentially unchanged apart from the appropriate map of parameters.  We expect a similar story holds for non-conformal gauge theories with a Coulomb branch, such as the toroidal $Dp$-branes for $p\ne 3$ described in section~\ref{sec:D-shells}, and suitable hyperbolic counterparts whose gravity solution is as far as we are aware unknown.  There is also the work of~\rcite{Henriksson:2019ifu,Henriksson:2021zei} on gauge theories dual to a deformation of $AdS_5\times T^{1,1}$, which again have an unstable Coulomb branch.  Finally, there are examples involving NS5-branes~\rcite{Itzhaki:2005zr}, which are currently under study~\rcite{MTW}.  In these last two examples, the Coulomb branch is destabilized by a suitable mass term rather than turning on spatial curvature in the gauge theory, but the effect is similar (and furthermore tuneable).

\vskip .5cm
Finally, it is perhaps
worth mentioning that the construction we converged upon has its ideological roots in non-critical string theory.  The fact that the 't Hooft large $N$ expansion is a sum over surfaces led to the idea that the path integral of 0+1d matrix quantum mechanics was a realization of 1+1d string theory (see~\rcite{Klebanov:1991qa,Ginsparg:1993is,Martinec:2004td} for reviews).  This and related constructions explored at the time are perhaps the first examples of gauge/gravity duality.  In the large $N$ limit, the matrix potential is $V(\sfX) = -\Tr[\sfX^2]$.  Gauging the $U(N)$ symmetry eliminates all of the non-abelian matrix dynamics; the only remnant of that structure is eigenvalue repulsion from the path integral measure, which turns the eigenvalues into fermions. 
Considering a ``ground state'' in which one populates the potential landscape with a Fermi sea of eigenvalues, there is an S-matrix that sends density perturbations in from infinity, which scatter off the side of the potential and back out, very much like the subcritical shells in hyperbolic $AdS$ depicted in figure~\ref{fig:hyperhistories}.  

The dual 1+1d string theory background has a linear dilaton, such that string dynamics is strongly coupled near the top of the matrix potential.  Revisiting this system in the light of gauge/gravity duality~\rcite{Douglas:2003up}, it was realized that the matrix eigenvalues can be thought of as D0-branes of 1+1d non-critical string theory.  And so the setup falls very much into the framework of gauge/gravity duality, where the collective dynamics of a gauge theory has a dual description as string dynamics in the background sourced by a condensate of D-branes.

Around the same time as the original developments, it was observed that there is an exact solution of the string theory beta function equations of motion describing a 1+1d dilatonic black hole~\rcite{Witten:1991yr}.  It seemed reasonable that there might be a non-perturbatively controlled setting for the analysis of black hole creation and evaporation.

But we now understand that the large entropy of black holes arises from non-abelian gauge dynamics.  The only effect of non-abelian structure here is to impose the Gauss law of the gauged $U(N)$ symmetry, which {\it eliminates} all the non-abelian structure of the matrix $\sfX$, leaving only its (anti)commuting eigenvalues.  The system doesn't have the degrees of freedom that are needed to describe black holes, and indeed there is no dynamics that traps the eigenvalue D0-branes near the top of the matrix potential.  From the bulk perspective, it turns out that the black hole solutions are non-normalizable excitations of the 1+1d linear dilaton vacuum~\rcite{Kazakov:2000pm,Giveon:2005mi}, and so are not part of the Hilbert space.

Thus, while this matrix quantum mechanics/2d string theory duality has an S-matrix, it does not have black holes as intermediate states.  However, this setup is very much the intellectual antecedent of the construction presented here.  What we have added are the black holes, which arise in the multi-matrix dynamics of D-branes if the matrix eigenvalues reach the strong-coupling region near the top of the unstable matrix potential.



\vskip 2cm

\noindent{\bf Acknowledgments:}
We thank 
Iosif Bena, 
Samir Mathur, 
David Turton, 
and 
Nicholas Warner 
for discussions.
The work of EJM is supported in part by DOE grant DE-SC0009924, as well as a France-UChicago FACCTS collaboration grant.  The hospitality of IPhT Saclay, the University of Southampton, and NCTS Taiwan during the writeup of our results is appreciated.

\vskip 2cm


\begin{appendix}

\section{Speculations on infall and the black hole interior}
\label{sec:interior}

An issue that our analysis has left largely unaddressed is the nature of the black hole interior, and the experience of the infalling observer.  That is largely because the effective action approach we have adopted describes only the black hole exterior, supplemented by the minimal set of ingredients needed to capture the black hole entropy and allow a unitary S-matrix.

In the effective supergravity theory itself, general coordinate invariance of the classical theory allows the description of the dynamics in a smooth foliation of the black hole spacetime (whose spatial hypersurfaces are so-called ``nice'' slices).  After the infalling brane shell has formed a horizon, the near-horizon region quickly settles down to the vacuum black hole solution.

An early proposal to reconcile this feature  with a unitary dynamics and resolve the information paradox was the notion of {\it black hole complementarity}~\rcite{Susskind:1993if} (see also~\rcite{Schoutens:1993hu,Schoutens:1994st}), which postulated that the infalling and asymptotic observers' descriptions of the state of the system were built out of non-commuting variables.  The outside observer's description ends at the horizon, or rather at a timelike surface slightly outside, dubbed the ``stretched'' horizon.  Black hole microstates in that description were supposed to live on the stretched horizon, absorbing and emitting energy and information according to the laws of quantum mechanics.  The bulk effective field theory employed in the present work is very much in the same spirit.  

Because quantum information cannot be duplicated, having that information stored in a set of degrees of freedom in the vicinity of the horizon seems to preclude the existence of a separate copy of that information in the infalling brane shell deep in the black hole interior on a nice slice.  Black hole complementarity proposed a way to rescue a smooth description for infalling observers by postulating that the infalling description employs a different basis of the Hilbert space that diagonalizes different observables, which don't commute with the observables of the asymptotic observer, which are diagonalized in a different basis.  The idea was thus that the information is not ``in two places at once'' because the notion of ``place'' is basis-dependent. 

The small corrections theorem of~\rcite{Mathur:2009hf} and the analysis of~\rcite{Almheiri:2012rt} argue that black hole complementarity is inconsistent with commonly held beliefs about how quantum gravity works~-- that the combination of the entanglement structure of the radiation, locality and unitarity is incompatible with a smooth horizon in the low-energy effective field theory.

A related issue is the question of what is meant by ``low energy'', and therefore which degrees of freedom are encompassed by the effective field theory, and which are integrated out.
Any proposed extension of the asymptotic observer's description of the black hole to encompass the black hole interior involves an analytic continuation of the vacuum exterior geometry past a coordinate singularity at the horizon.  The low-energy effective field theory of infall involves high energy data according to asymptotic observers, data that in principle lies outside the scope of their effective bulk description.%
\footnote{The point being made here is not that there are trans-Planckian effects in the Hawking process that somehow call into question the use of effective field theory near the horizon, as is sometimes claimed on the basis of a misunderstanding of how vacuum fluctuations gradually become on-shell particles as they propagate away from the near-horizon region.  Rather, it is that any attempt to reconstruct near-horizon physics using the asymptotic observer's description requires consideration of degrees of freedom that have been integrated out of their low-energy effective field theory. }  

There is also room for an order of limits issue to obstruct such an analytic continuation.  If one takes $N\to\infty$ first, the classical black hole geometry is exact, and then one can freely analytically continue it past the horizon (though of course, there is then no information paradox to discuss, because the black hole does not radiate).  On the other hand, if one keeps $N$ finite, and attempts to analytically continue the properties of any given microstate, the blueshift involved in that continuation amplifies arbitrarily the tiny disparities in near-horizon structure between the classical near-horizon black hole geometry and that of the given microstate, potentially obstructing the existence of a smooth low-energy near-vacuum state in the black hole interior at finite $N$.  An example of such blueshift effects in a closely related context is analyzed in~\rcite{Martinec:2020cml}.

A novel proposal for a description of the interior was given in~\rcite{Horowitz:2009wm}.
The hyperbolic D3-brane geometry for $\mu=0$  is a hyperbolic slicing of past light cones  in the Poincaré patch of $AdS_5$.  The coordinate transformation 
\be
t_p = -\frac{r\,e^{-t}}{(r^2-1)^{1/2}}
~~,~~~~
r_p = (r^2-1)^\half\,e^t
\ee
maps the static exterior geometry to a Milne-type parametrization of the Poincar\'e AdS metric, which continues smoothly into the black hole interior and all the way to the singularity
\be
\label{poincare map}
ds^2 = \ell^2\bigg[ r_p^2\Big(-dt_p^2+t_p^2\,d\Sigma_3^2\big)+\frac{dr_p^2}{r_p^2} \bigg] ~.
\ee
Setting $\mu=0$ in~\eqref{mu and v} and~\eqref{vels} and integrating to find the trajectory of the infalling shell, one finds that it follows the geodesic
\be
r_p = {\it const.} ~,
\ee
reproducing the analysis of~\rcite{Horowitz:2009wm}.  These authors proposed that the gauge theory on hyperbolic Milne space is dual to the bulk described in the above Poincar\'e slicing of the $\mu=0$ geometry, and thus provides a smooth description of the black hole interior. 

While this is a natural guess, and the two gauge theories are naively related by a (singular) conformal transformation (in particular, time in one is related to time in the other by the $r\to\infty$ limit of~\eqref{poincare map}, $t_p = -e^{-t}$), it is not obvious that this is the right map, especially when it comes to the properties of the out state.
If there is such a map, then the asymptotic late-time state consisting of an outgoing flux of branes must be somehow encoded in the high-energy properties of the state of the gauge theory at a Milne singularity.  

Gauge theory on Milne space is {\it not} amenable to an effective field theory description of the bulk.  Rather, the gauge theory dynamics accesses arbitrarily high frequency excitations as the Milne singularity is approached.  
In particular, there are large fluctuations of the eigenvalues of the SYM matrices, which would have to encode the random and chaotic release of branes in the S-matrix out state.
After the shell crosses the horizon, 
the proper time to the singularity is shorter than the light-crossing time on $\Sigma_3$.  Every halving of the time to the singularity doubles the cutoff on modes whose vacuum fluctuations are growing incoherently.
Again, this is true for all $N^2$ degrees of freedom in the gauge theory, so these virtual fluctuations encompass all sorts of stringy excitations, not just the $N^0$ collective modes that are dual to supergravity.  
Therefore, in this particular proposed dual description of the black hole interior, there is no bulk effective field theory of the black hole interior which involves only supergravity rather than all of string theory.  One then has a rather pyrrhic victory of complementarity~-- a description of the black hole interior that involves all $N^2$ degrees of freedom of the gauge theory, which indeed do not commute with those describing the branes well out on their Coulomb branch, but at the same time do not have any approximation in which the black hole interior is the smooth vacuum spacetime predicted by general relativity.

\end{appendix}



\newpage
\bibliographystyle{JHEP}
\bibliography{fivebranes}

\end{document}
